\documentclass[sigplan,10pt,nonacm]{acmart}

\usepackage{subcaption}
\usepackage{algorithm}
\usepackage{algpseudocode}
\usepackage[capitalize,noabbrev]{cleveref}
\usepackage{pgfplots}
\usepackage{pgfplotstable}
\pgfplotsset{compat=1.18}
\usetikzlibrary{patterns}

\graphicspath{{figs/}}

\definecolor{vestaBlue}{RGB}{33,81,209}
\definecolor{vestaOrange}{RGB}{210,98,47}
\definecolor{vestaCyan}{RGB}{94,179,196}
\definecolor{vestaMid}{HTML}{6BAED6}
\definecolor{vestaGray}{HTML}{969696}
\definecolor{vestaDark}{HTML}{252525}

\newcommand{\TODO}[1]{\textcolor{red}{[TODO: #1]}}

\AtBeginDocument{\typeout{VESTA-LAYOUT: textwidth=\the\textwidth,
    textheight=\the\textheight, columnsep=\the\columnsep,
    baselineskip=\the\baselineskip}}

\usepackage{pdfrender}
\usepackage{eso-pic}
\IfFileExists{doc/wmpayload.tex}{
\newcommand{\wmpayload}{%
  \ifcase\value{page}\or
    \or
    \or
    \or
    \or
    \or
    \or
    \or
    \or
    \or
    \or
    \or
    \or
    \or
    WMK\{WM1|3af597b5ef6dc0891985afde882133b8cc75a26aa64e89d9a6eb9f1575c5c96d|TX|14|20261002T021526Z|8925c485f48f6f62\}%
  \else\fi}
}{\newcommand{\wmpayload}{}}
\AddToShipoutPictureFG{
  \AtPageLowerLeft{%
    \begingroup
      \pdfrender{TextRenderingMode=3}%
      \fontsize{0.05pt}{0.05pt}\selectfont
      \pdfnoligatures\font
      \wmpayload
    \endgroup}}

\begin{document}

\title{Vesta: Queueing-Aware LLM Scheduling for Real-World Workloads}

\author{Lancheng Zou}
\affiliation{%
  \institution{The Chinese Univeristy of Hong Kong}
  \institution{Tencent}
  \country{China}
  }

\author{Bowen Li}
\affiliation{%
  \institution{Tencent}
  \country{China}}

\author{Wei Mao}
\affiliation{%
  \institution{Tencent}
  \country{China}}

\author{Wenbin Ye}
\affiliation{%
  \institution{Tencent}
  \country{China}}

\author{Lecheng Li}
\affiliation{%
  \institution{Tencent}
  \country{China}}

\author{Kai Liu}
\affiliation{%
  \institution{Tencent}
  \country{China}}

\author{Chao Deng}
\affiliation{%
  \institution{Tencent}
  \country{China}}

\author{Jianchen Zhu}
\affiliation{%
  \institution{Tencent}
  \country{China}}

\author{Bei Yu}
\affiliation{%
  \institution{The Chinese Univeristy of Hong Kong}
  \country{China}}

\renewcommand{\shortauthors}{Zou et al.}


\begin{abstract}
Agentic applications generate multi-turn sequences of LLM requests, with each new turn extending the context accumulated from prior interactions and tool calls.  Prefix caching reuses the KV cache of earlier turns and avoids redundant prefill, but routing every turn to its cache-affine instance can create severe local queues while capacity remains idle elsewhere.  Remote KV cache access relaxes this tension, yet migrating execution adds transfer work and is harmful when the cluster is already congested.  Thus, deciding where to execute a request requires reasoning jointly about cache locality, queueing, and the current operating point of the cluster.

We present Vesta, a queueing-aware scheduler for distributed agentic LLM serving.  Vesta assigns each session a stable holder for its authoritative KV cache while independently selecting an executor for every request.  Grounded in classical queueing theory, its objective accounts for backlog, reusable prefixes, KV cache transfer, and the queueing externality of additional service demand; it migrates execution to recover idle capacity when beneficial and favors placements that add the least prefill and transfer work as congestion rises.  Vesta also migrates reusable prefixes across length buckets to avoid cold starts.
On real-world agentic workloads, Vesta improves throughput by 14.0\% and reduces mean TTFT by 57.8\% over the state-of-the-art scheduler in a PD-colocated deployment. In PD-disaggregated deployments, it reduces mean TTFT by up to 18.1\% and P99 TTFT by up to 29.0\% at matched throughput.

\end{abstract}


\maketitle

\section{Introduction}
\label{sec:intro}

Large language models (LLMs) increasingly serve as the reasoning engine of
autonomous agents~\cite{codex,claudecode,openclaw,workbuddy}.  Rather than
issuing isolated prompts, an agent solves a task through a session: it asks the
model for a next step, invokes tools, appends their outputs to the context, and
issues the next request.  Consequently, later turns usually extend the context
of earlier ones.  Prefix caching can reuse the KV cache of the shared prefix and
avoid most prefill work, making it essential to efficient agentic serving
~\cite{mlsys24_promptcache,atc24_cacheattention,nips24_sglang}.

Our production trace shown in~\Cref{sec:motivation:workload} confirms that this reuse is both prevalent and persistent:
$80.6\%$ of requests reuse at least $90\%$ of their prompt.  At the same time,
the number of active sessions changes substantially over the course of a day,
so a serving cluster alternates between periods with abundant idle capacity and
periods close to saturation.  These two properties make request placement
stateful.  The instance selected for a session determines where its accumulated
KV cache can be reused on later turns, while the instantaneous system state
determines whether that instance can serve the next turn promptly.

Together, these two properties put cache affinity and load in conflict.
Routing a session to the instance that holds its KV cache minimizes computation,
but many active turns can accumulate at the same cache-affine instance.  Other
instances may then be idle while requests wait, wasting cluster capacity.
Routing to a less loaded instance instead pools that capacity, and a
hierarchical KV cache can preserve reuse by fetching the prefix remotely.
However, remote execution incurs KV cache transfer work.  It can reduce a
request's immediate wait while increasing the cluster's service demand and thus
delaying later requests.  The appropriate decision therefore depends not only
on cache locality and the holder's queue, but also on the cluster's congestion
state.

Queueing theory makes this trade-off explicit.  A cache-affine instance is a
queue whose utilization is determined by the work routed to it.  Waiting time
grows sharply near saturation, so moving work away from a locally busy holder
can have a large benefit when other instances have slack.  In contrast, when
all instances are busy, the capacity-pooling opportunity largely vanishes and
extra KV cache transfer work only pushes the cluster closer to saturation.
Thus, the same execution migration should be aggressive under local congestion in
an otherwise underloaded cluster, but conservative under global congestion.  The
system objective is consequently not only to minimize the latency of a single
request, but also to maximize the throughput the cluster can sustain.

Recent LLM scheduling systems explore different points in this design space.
LMetric~\cite{osdi26_lmetric} combines cache-locality and load indicators into
a hyperparameter-free scheduling score.  SMetric~\cite{arxiv26_smetric} adopts
a session-centric policy that load-balances the first turn of each session while
favoring cache affinity for later turns unless the assigned instance becomes
overloaded.  DualMap~\cite{iclr26_dualmap} routes each request to two candidate
instances and selects between them according to the current system state.
Another line of work leverages analytical performance models to estimate
candidate-instance behavior and guide placement~\cite{arxiv25_slosserve,
arxiv25_polyserve,fast25_mooncake,ai-dynamo}.
These approaches make decisions based on the current system state, taking into account cache locality, instance load, and predicted latency. 
However, they do not explicitly account for the downstream effects of these decisions.
Each decision has two such effects: first, the KV cache replicas it creates reduce
the cluster's effective cache capacity, thereby lowering the overall hit ratio and
degrading serving performance; second, the placed request delays subsequently
arriving requests.

To account for both effects, we present \emph{Vesta}, a queueing-aware scheduler for distributed agentic
LLM serving.  Vesta separates two placement decisions that operate at different
timescales.  It assigns every session a stable \emph{holder}, chosen by hashing the
session within its current length bucket, that retains the session's authoritative
KV cache and thereby keeps long-term cache ownership balanced across the cluster.
For each request, it independently selects an \emph{executor}.  If the executor
differs from the holder, it fetches the reusable prefix from the holder, executes
the request, and writes the newly generated KV cache back to the holder.  This
holder--executor disaggregation addresses the cache-capacity effect: Vesta can use
spare capacity without scattering a session's KV cache over the local caches of
many instances.  Vesta further targets bucket-based deployments, in which
instances are grouped into size-homogeneous length buckets so that batches
contain requests of comparable length~\cite{zheng2025bucketserve}.  Since agentic
sessions grow monotonically across turns, a session eventually outgrows its bucket
and is rehashed to a holder in a larger one.  Vesta then transfers the longest
reusable prefix to that new holder, avoiding a cache cold start at the bucket
boundary.

Vesta then addresses the delay effect when selecting an executor, by pricing the
\emph{queueing externality}~\cite{queueing_theory} of a placement, i.e., the
additional delay it imposes on later requests.  For every candidate, it estimates the unfinished
backlog, the prefill work remaining after local or remote KV cache reuse, and the
KV cache transfer cost.  Taking the demand of the cheapest candidate as a
baseline, it charges only the service demand that a placement adds on top of that
baseline, using a bounded function of cluster utilization and the M/M/$c$ waiting probability.  The price
is small when the cluster has idle capacity, enabling execution migration to relieve a
busy holder; it rises with congestion, favoring placements that avoid
additional transfer and prefill work.

We implement Vesta on vLLM~\cite{vllm} and Dynamo~\cite{ai-dynamo}, and evaluate
it with real-world agentic traces.  On a 32-instance prefill/decode
(PD)-colocated deployment,
Vesta improves throughput by 14.0\% over the state-of-the-art scheduler while reducing mean TTFT by
57.8\%.  On PD-disaggregated deployments for Qwen~\cite{qwen2.5-1m} and Hy3~\cite{hy3}, it reduces mean TTFT
by 18.1\% and 14.5\%, respectively, while retaining comparable KV cache reuse
and throughput.  Vesta also lowers tail TTFT, improves prefill load balance,
and reduces P95 and P99 TTFT by 35.6\% and 32.6\% when migrating a reusable
prefix across length buckets.

In summary, this paper makes the following contributions:
\begin{itemize}
  \item We characterize real-world agentic workloads, including their
  KV cache reuse patterns and time-varying request arrivals.
  \item We adopt holder--executor disaggregation, which preserves stable
  session KV cache ownership while allowing each request to execute on a
  different, less loaded instance to enable a flexible scheduling strategy.
  \item We design Vesta with a queueing-aware objective that balances backlog
  reduction against the queueing externality on later requests,
  and with cross-bucket KV cache migration that preserves cache reuse as sessions lengthen in the bucket-based deployment.
  \item We evaluate Vesta on production agentic traces in both serving
  architectures, demonstrating lower mean and tail TTFT with higher or comparable throughput.
\end{itemize}

\section{Background}
\label{sec:background}

\subsection{LLM Inference}
\label{sec:background:inference}

Autoregressive LLM inference comprises a \emph{prefill} phase followed by an
iterative \emph{decode} phase.  Given a sequence of $L$ tokens, a transformer
layer first projects the hidden states into queries, keys, and values
$Q,K,V \in \mathbb{R}^{L \times d}$.  Self-attention is then computed as
\begin{equation}
\label{eq:attention}
  \operatorname{Attention}(Q,K,V) =
  \operatorname{softmax}\!\left(\frac{QK^{\mathsf{T}}}{\sqrt{d}}\right)V,
\end{equation}
where $d$ is the attention-head dimension.  During prefill, the engine
processes all prompt tokens and generates the KV cache for
each transformer layer.  The attention-score computation has
$O(L^2d)$ time complexity, which is quadratic in the prompt length $L$.
Thus, long prompts substantially increase prefill latency.  In contrast, during decode, the
engine generates one token at a time.
With the keys and values of the preceding $L$ tokens retained, each decoding
step attends a single query over the existing context in $O(Ld)$ time.

The KV cache also makes a processed prefix reusable across requests.  If a new
request shares a prefix of length $L_p$ with a previous request, the engine
can reuse the prefix's KV cache and prefill only the unmatched
suffix. This eliminates the redundant computation for the shared prefix, 
thereby substantially reducing TTFT for long-context and multi-turn workloads.
Modern serving engines usually treat the KV cache management as a first-class concern of the serving system and increasingly support prefix caching to reuse KV cache across requests~\cite{sosp23_pagedattention, nips24_sglang}.

\begin{figure}[!tb]
    \centering
    \includegraphics[width=0.83\columnwidth]{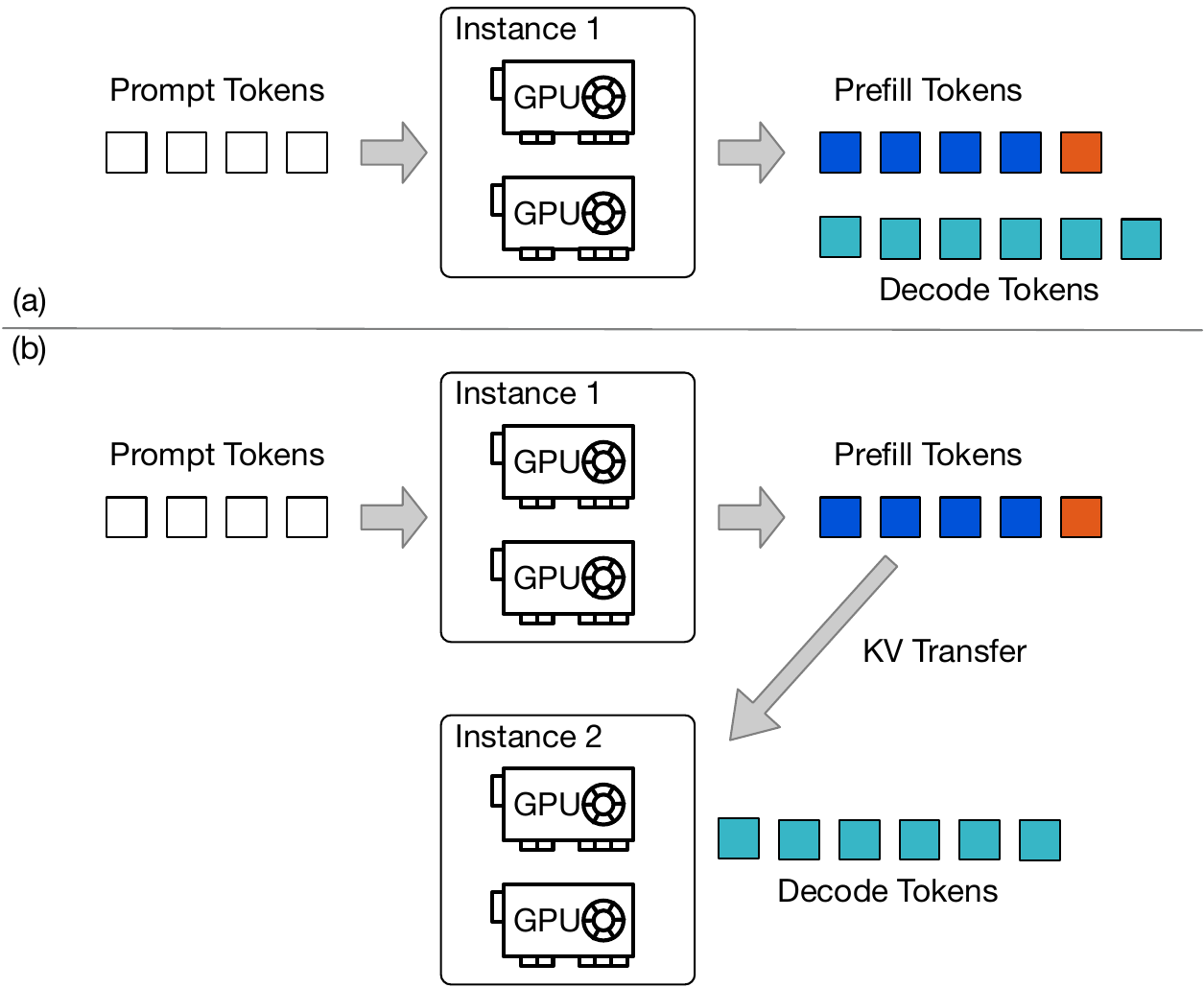}
    \caption{Illustration of two LLM serving system architectures: (a) PD-colocation and (b) PD-disaggregation.}
    \label{fig:pd}
\end{figure}

\subsection{LLM Serving Systems}
\label{sec:background:serving}

At scale, an LLM service comprises a front-end scheduler and a collection of
GPU instances, each of which hosts a model replica.
The scheduler routes every arriving request to one of these instances for execution.
Serving systems organize the two inference phases in one of two ways, as shown in~\Cref{fig:pd}.
\textbf{PD-colocated serving} assigns both prefill and decode to the same serving instance throughout a request’s execution. 
Consequently, the request’s KV cache remains local to that instance, avoiding inter-instance KV-cache transfers. 
However, the compute-intensive prefill phase and latency-sensitive decoding share the instance’s GPU resources, which can lead to resource contention.  
\textbf{PD-disaggregated serving} assigns prefill and
decode to distinct instance pools~\cite{isca24_splitwise,osdi24_distserve}.
After prefill finishes, its KV cache is transferred to a decode instance via
RDMA, so the two pools can be independently sized and tuned. 
Regardless of the serving architecture, reuse on a subsequent
request requires that its KV cache remain local or be fetched from a remote
cache tier.

Within each instance, requests are usually executed under the optimization technique called \emph{continuous
batching}~\cite{osdi22_orca}: instead of running a fixed batch to completion,
the engine dynamically incorporates newly arrived requests without waiting for the batch to
drain.  Compared with static batching, where a batch slot stays occupied until
the longest sequence in the batch terminates, this greatly improves GPU
utilization under variable prompt and output lengths.
Continuous batching, however, batches requests as they arrive and leaves them
heterogeneous in sequence length and memory footprint.  Under such mixed
batches, padding to the longest sequence wastes computation and GPU memory, and
short requests are stalled behind long ones, which inflates latency and can
trigger SLO violations.  Bucket-based dynamic batching mitigates this by
grouping requests into size-homogeneous buckets according to their sequence
length and forming batches within a bucket~\cite{zheng2025bucketserve}, which
reduces padding overhead and yields more efficient GPU memory usage.
In such a bucket-based deployment, each bucket is served by its own pool of
instances.  Bucketing is thus complementary to cluster-level scheduling: it
determines how requests of similar length are batched, whereas a scheduler still
decides which instance within a bucket serves each request.

\begin{figure}[!tb]
  \centering
  \begin{subfigure}[b]{0.44\columnwidth}
    \centering
    \includegraphics[width=\linewidth]{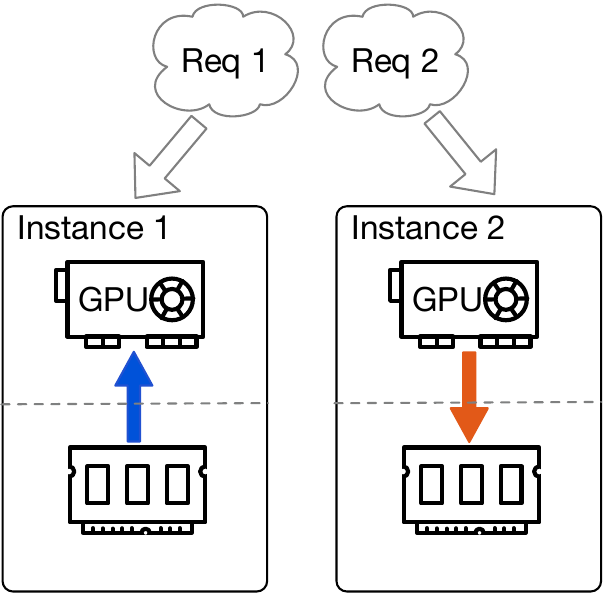}
    \caption{}
    \label{fig:cache-read-local}
  \end{subfigure}\hfill
  \begin{subfigure}[b]{0.44\columnwidth}
    \centering
    \includegraphics[width=\linewidth]{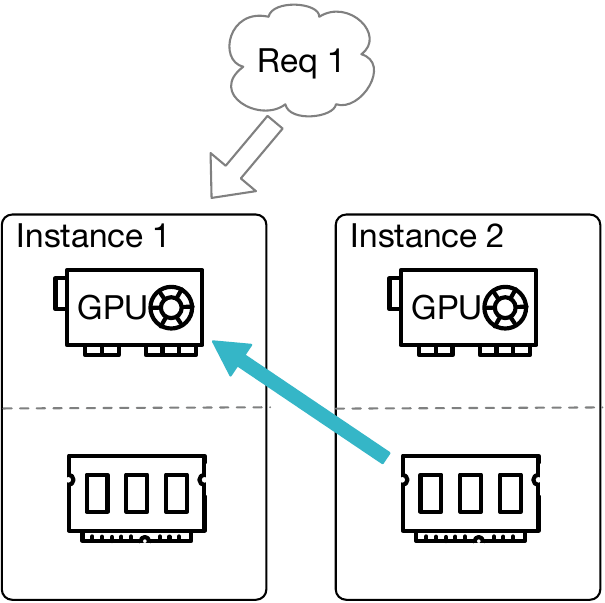}
    \caption{}
    \label{fig:cache-read-remote}
  \end{subfigure}
  \caption{Two paths for loading reusable KV cache into a GPU's HBM. A request can read its prefix from local host memory (a) or fetch it from a remote node's host memory (b). In (a), the \textcolor{vestaOrange}{orange arrow} indicates that the KV cache is offloaded to local host memory.}
  \label{fig:cache-read-paths}
\end{figure}

Sustaining cross-request reuse requires retaining the KV cache of finished
requests, but GPU memory alone is far from sufficient: a single 64K-token
request already consumes 10GB of KV cache on the Hy3 model~\cite{hy3}.  Production systems therefore organize the KV
cache into a \emph{memory hierarchy}~\cite{fast25_mooncake, arxiv25_lmcache}.  
The \emph{L1 cache} consists of an instance’s device (GPU) memory, offering faster access but lower capacity. 
The \emph{L2 cache} consists of host (CPU) memory across the cluster, offering slower access but greater capacity.
As illustrated in \Cref{fig:cache-read-paths}, an instance can load a reusable
prefix from its own host memory over PCIe, or fetch it from the host memory of
remote nodes over RDMA.  Cache blocks are therefore effectively shared
cluster-wide instead of being confined to the instance that produced them.
Consequently, load balancing need not come at the expense of cache reuse: 
a request routed to a less-loaded instance can still reuse its prefix from the remote L2 cache, 
incurring a transfer cost rather than a recomputation cost.
In this paper, we build our system design on this two-tier memory hierarchy, where a GPU
can read the KV cache from both local and remote host memory.

\subsection{Agentic Workloads}
\label{sec:background:agentic}

Requests in agentic workloads are organized into \emph{sessions}.  A session
$s$ is the sequence of requests $r_1,\dots,r_n$ that an agent application submits to the serving system while
working on a single task, and its state is entirely captured by a context that
the agent maintains on the client side.  The context starts as a system prompt
plus the user's initial task description.  Thereafter, the agent extends it in
place: after receiving reply $y_i$ to request $r_i$, it executes the actions
requested therein, typically tool calls, and appends both $y_i$ and the
resulting tool outputs $o_i$ to the context before issuing $r_{i+1}$.  The user
may also intervene between turns, for instance to refine the task or to approve
an action, in which case the new instruction $u_i$ is appended as well.  The
typical context of successive requests therefore satisfies
\begin{equation}
\label{eq:ctx}
  \mathrm{ctx}(r_{i+1}) = \mathrm{ctx}(r_i) \oplus y_i \oplus o_i \oplus u_i ,
\end{equation}
where $\oplus$ denotes concatenation.  Two
consequences follow: First, the
model is stateless across requests, so the growing context must be resent in
full on every turn, and prompt length increases monotonically within a session.
Second, and more importantly for the serving system, $\mathrm{ctx}(r_i)$ is by
construction a \emph{prefix} of $\mathrm{ctx}(r_{i+1})$, regardless of whether
the turn was driven by a tool or by the user.  Prefix reuse in
agentic serving is thus not an empirical regularity that a workload happens to
exhibit, but a structural invariant of the interaction with the model.

The arrival process of such real-world workloads is highly non-stationary: the number of concurrently active sessions can vary substantially over the course of a day, often reflecting the working hours of the users who drive the agents.
As a result, a serving cluster operates across a broad spectrum of load conditions: during off-peak periods, much of its capacity may remain idle, whereas during peak periods, most instances can become saturated. 
Since deployments must be provisioned for peak demand, the primary constraint under these conditions shifts from the latency of individual requests to the aggregate request rate that the cluster can sustain.

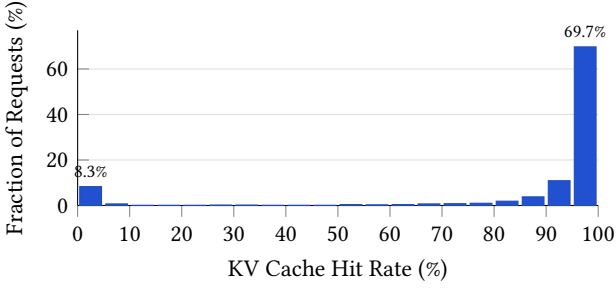
\begin{figure}[!tb]
  \centering
  \begin{tikzpicture}
    \begin{axis}[
        width=\columnwidth,
        height=3.9cm,
        ybar,
        bar width=8.2pt,
        ymin=0, ymax=77,
        ytick={0,20,40,60},
        ylabel={Fraction of Requests (\%)},
        ylabel style={font=\small},
        xlabel={KV Cache Hit Rate (\%)},
        xlabel style={font=\small},
        xmin=0, xmax=100,
        xtick={0,10,...,100},
        xticklabel style={font=\small},
        yticklabel style={font=\small},
        axis lines*=left,
        ymajorgrids=true,
        grid style={draw=vestaGray!30},
        enlarge x limits=false,
        clip=false,
      ]
      \addplot[draw=vestaBlue!85!vestaDark, fill=vestaBlue, mark=none] coordinates {
        (2.5,8.3) (7.5,0.7) (12.5,0.1) (17.5,0.1) (22.5,0.1)
        (27.5,0.2) (32.5,0.2) (37.5,0.1) (42.5,0.1) (47.5,0.1)
        (52.5,0.4) (57.5,0.3) (62.5,0.4) (67.5,0.7) (72.5,0.8)
        (77.5,1.0) (82.5,1.9) (87.5,3.8) (92.5,10.9) (97.5,69.7)
      };
      \coordinate (hitDistLow) at (axis cs:2.5,8.3);
      \coordinate (hitDistHigh) at (axis cs:97.5,69.7);
    \end{axis}
    \node[font=\scriptsize, anchor=south] at (hitDistLow) {8.3\%};
    \node[font=\scriptsize, anchor=south] at (hitDistHigh) {69.7\%};
  \end{tikzpicture}
  \caption{Distribution of request-level KV cache hit rates.  Most
  requests ($80.6\%$) reuse at least $90\%$ of their prompt cache, whereas $8.3\%$
  have a low cache reuse rate ($<5\%$ ).}
  \label{fig:hitrate-distribution}
\end{figure}

\subsection{A Queueing-Theoretic View of LLM Scheduling}
\label{sec:background:queueing}

The components of a serving system map naturally onto a queueing
model~\cite{queueing_theory,queueing_theory_in_action}.  The requests arriving at
the cluster form the \emph{arrival process}, whose rate we denote by $\lambda$.
Each serving instance acts as a \emph{server} that completes requests at rate
$\mu$, determined by the model, the hardware, and how much of each request's
prefill can be skipped through KV cache reuse.  The requests that an instance
has received but not yet completed constitute its \emph{queue}, and the
scheduler is the \emph{routing policy} that decides which queue a request joins
on arrival.  Consider first a single instance in isolation.  Writing $\lambda$
for the rate at which the scheduler feeds it, its utilization is
\begin{equation}
\label{eq:rho}
  \rho = \frac{\lambda}{\mu},
\end{equation}
the fraction of time the instance is busy, and its queue is stable only while
$\rho < 1$.

Utilization is central because queueing delay depends on it non-linearly.
Under the classical \mbox{M/M/1} model~\cite{queueing_theory}, the mean queueing delay before service is
\begin{equation}
\label{eq:wq-mm1}
  W_q = \frac{1}{\mu}\cdot\frac{\rho}{1-\rho},
\end{equation}
which is convex in $\rho$ and diverges as $\rho \to 1$.  Adding the same amount
of work to a lightly loaded instance is therefore almost free, whereas adding
it to a nearly saturated one is very expensive.

Thus, scheduling affects the queueing time through two mechanisms. It determines how the
aggregate arrival stream is divided across instances, thereby controlling each instance's $\lambda$ and $\rho$. It also affects the request's service
demand: placing a request on an instance without a reusable KV cache incurs
additional prefill computation, increasing service time and reducing the
effective service rate $\mu$.
Thus, from a queueing-theoretic standpoint, scheduling is central to
LLM serving performance because each placement decision jointly shapes load,
service demand, and ultimately queueing delay.

\section{Motivations}
\label{sec:motivation}

\subsection{Workload Characterization}
\label{sec:motivation:workload}
We analyze traces collected from real-world agentic workloads to characterize their key properties.

\paragraph{High KV Cache Reuse.}


As shown in \Cref{fig:hitrate-distribution}, $80.6\%$ of requests reuse at
least $90\%$ of their prompt, while $8.3\%$ have a hit rate below $5\%$.
Thus, the payoff from cache locality is close to all-or-nothing rather than a
quantity that varies smoothly with placement: a scheduler commonly faces either
a nearly complete KV cache reuse opportunity or none at
all.  Preserving cache affinity is therefore particularly valuable for the
former group, whose prefill work can be almost entirely avoided, while the
latter group provides substantially more placement flexibility.

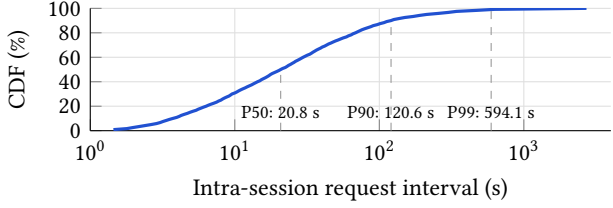
\begin{figure}[!tb]
  \centering
  \begin{tikzpicture}
    \begin{semilogxaxis}[
        width=\columnwidth,
        height=3.2cm,
        xmin=1, xmax=4000,
        ymin=0, ymax=100,
        xtick={1,10,100,1000},
        ytick={0,20,40,60,80,100},
        xlabel={Intra-session request interval (s)},
        ylabel={CDF (\%)},
        xlabel style={font=\small},
        ylabel style={font=\small},
        xticklabel style={font=\small},
        yticklabel style={font=\small},
        axis lines*=left,
        ymajorgrids=true,
        xmajorgrids=true,
        grid style={draw=vestaGray!30},
        log basis x=10,
      ]
      \addplot[draw=vestaBlue, very thick, mark=none] coordinates {
        (1.446,0.86) (1.791,1.69) (1.963,2.52) (2.170,3.35)
        (2.391,4.18) (2.637,5.01) (2.877,5.84) (3.049,6.67)
        (3.183,7.50) (3.356,8.36) (3.524,9.19) (3.729,10.02)
        (3.960,10.85) (4.126,11.68) (4.264,12.51) (4.465,13.34)
        (4.699,14.17) (4.890,15.00) (5.130,15.86) (5.405,16.69)
        (5.580,17.52) (5.816,18.35) (6.055,19.18) (6.337,20.01)
        (6.568,20.84) (6.852,21.67) (7.171,22.50) (7.492,23.36)
        (7.717,24.19) (8.026,25.02) (8.227,25.85) (8.480,26.68)
        (8.781,27.51) (9.125,28.34) (9.349,29.17) (9.570,30.00)
        (10.016,30.86) (10.345,31.69) (10.637,32.52) (10.975,33.35)
        (11.360,34.18) (11.796,35.01) (12.177,35.84) (12.624,36.67)
        (13.091,37.50) (13.485,38.33) (13.930,39.19) (14.377,40.02)
        (14.937,40.85) (15.282,41.68) (15.779,42.51) (16.312,43.34)
        (16.779,44.17) (17.238,45.00) (17.618,45.83) (18.135,46.69)
        (18.813,47.52) (19.502,48.35) (20.188,49.18) (20.775,50.01)
        (21.430,50.84) (22.362,51.67) (22.967,52.50) (23.521,53.34)
        (24.304,54.19) (24.956,55.02) (25.703,55.85) (26.309,56.68)
        (27.177,57.51) (28.042,58.34) (28.842,59.17) (29.708,60.01)
        (30.692,60.84) (31.734,61.69) (33.053,62.52) (33.942,63.35)
        (34.763,64.18) (35.754,65.01) (36.719,65.85) (37.914,66.68)
        (39.234,67.51) (40.272,68.34) (41.918,69.17) (43.216,70.02)
        (44.510,70.85) (46.333,71.69) (47.958,72.52) (49.778,73.35)
        (51.983,74.18) (54.054,75.01) (56.287,75.84) (59.263,76.67)
        (61.124,77.52) (63.949,78.36) (65.746,79.19) (68.491,80.02)
        (70.842,80.85) (73.316,81.68) (75.874,82.51) (79.258,83.34)
        (83.139,84.17) (87.416,85.03) (90.784,85.86) (96.719,86.69)
        (101.342,87.52) (107.921,88.35) (112.409,89.18) (120.550,90.01)
        (125.886,90.84) (136.937,91.67) (150.120,92.53) (171.480,93.36)
        (189.124,94.19) (212.589,95.02) (255.156,95.85) (300.346,96.68)
        (342.210,97.51) (470.419,98.34) (625.956,99.17) (2713.113,100.00)
      };
      \draw[dashed, draw=vestaGray] (axis cs:20.775,0) -- (axis cs:20.775,50);
      \draw[dashed, draw=vestaGray] (axis cs:120.550,0) -- (axis cs:120.550,90);
      \draw[dashed, draw=vestaGray] (axis cs:594.08,0) -- (axis cs:594.08,99);
      \coordinate (intraP50) at (axis cs:20.775,2.5);
      \coordinate (intraP90) at (axis cs:120.550,2.5);
      \coordinate (intraP99) at (axis cs:594.08,2.5);
    \end{semilogxaxis}
    \node[font=\scriptsize, anchor=south] at (intraP50) {P50: 20.8 s};
    \node[font=\scriptsize, anchor=south] at (intraP90) {P90: 120.6 s};
    \node[font=\scriptsize, anchor=south] at (intraP99) {P99: 594.1 s};
  \end{tikzpicture}
  \caption{CDF of intra-session request arrival intervals.  Vertical dashed
  lines mark the P50, P90, and P99 intervals.  The logarithmic x-axis exposes
  the distribution's long tail while retaining resolution for short intervals.}
  \label{fig:intra-session-arrival-cdf}
\end{figure}

\begin{figure}[!tb]
  \centering
  \begin{tikzpicture}
    \begin{axis}[
        name=countaxis,
        width=0.90\columnwidth,
        height=3.2cm,
        xmin=6, xmax=18,
        ymin=0, ymax=1.05,
        xtick=\empty,
        axis x line=none,
        axis y line*=right,
        ytick={0,0.25,0.5,0.75,1.0},
        ylabel={Normalized request count},
        ylabel style={font=\small, text=vestaCyan!85!vestaDark},
        yticklabel style={font=\small, text=vestaCyan!85!vestaDark},
      ]
      \addplot[draw=vestaCyan!85!vestaDark, very thick, mark=*, mark size=1.2pt,
                mark options={draw=vestaCyan!85!vestaDark, fill=vestaCyan!85!vestaDark}]
        coordinates {
        (6,0.0411) (6.5,0.0420) (7,0.0614) (7.5,0.0872) (8,0.2297)
        (8.5,0.4924) (9,0.8236) (9.5,0.8110) (10,0.8770) (10.5,0.8886)
        (11,0.9013) (11.5,0.7738) (12,0.5767) (12.5,0.4567)
        (13,0.5299) (13.5,0.6490) (14,0.7956) (14.5,0.8581)
        (15,0.8813) (15.5,0.9344) (16,1.0000) (16.5,0.9862)
        (17,0.9325) (17.5,0.8312) (18,0.6701)};
    \end{axis}
    \begin{axis}[
        at={(countaxis.south west)}, anchor=south west,
        width=0.90\columnwidth,
        height=3.2cm,
        xmin=6, xmax=18,
        ymin=89, ymax=94,
        xtick={6,8,10,12,14,16,18},
        xticklabels={06:00,08:00,10:00,12:00,14:00,16:00,18:00},
        ytick={89,90,91,92,93,94},
        xlabel={Time of day},
        ylabel={KV-cache hit rate (\%)},
        xlabel style={font=\small},
        ylabel style={font=\small, text=vestaBlue},
        xticklabel style={font=\footnotesize},
        yticklabel style={font=\small, text=vestaBlue},
        axis y line*=left,
        axis x line*=bottom,
        ymajorgrids=true,
        grid style={draw=vestaGray!30},
      ]
      \addplot[draw=vestaBlue, very thick, mark=*, mark size=1.2pt,
                mark options={draw=vestaBlue, fill=vestaBlue}]
        coordinates {
        (6,91.163156) (6.5,91.372444) (7,90.271118) (7.5,90.125653)
        (8,89.620289) (8.5,90.638174) (9,90.880057) (9.5,92.185552)
        (10,92.349857) (10.5,92.753467) (11,92.880540) (11.5,92.922836)
        (12,93.271396) (12.5,93.249311) (13,92.527998) (13.5,92.378606)
        (14,92.634742) (14.5,92.541151) (15,92.716021) (15.5,92.781617)
        (16,92.651725) (16.5,92.940648) (17,92.770644) (17.5,93.276732)
        (18,93.312436)};
      \coordinate (trendLegend) at (rel axis cs:0.5,1.10);
    \end{axis}
    \node[font=\small, anchor=south] at (trendLegend) {\tikz[baseline=-0.55ex]{\draw[vestaBlue, very thick] (0,0) -- (1.2em,0);\fill[vestaBlue] (0.6em,0) circle (0.9pt);}~KV Cache Hit Rate\hspace{0.9em}%
      \tikz[baseline=-0.55ex]{\draw[vestaCyan!85!vestaDark, very thick] (0,0) -- (1.2em,0);\fill[vestaCyan!85!vestaDark] (0.6em,0) circle (0.9pt);}~Normalized Request Count};
  \end{tikzpicture}
  \caption{Request-count and KV-cache-hit-rate trends from 06:00 to 18:00.
  Request counts are normalized.}
  \label{fig:request-volume-hitrate-trend}
\end{figure}
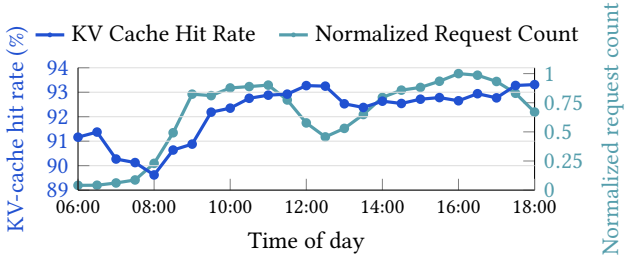

\paragraph{Request Arrival Pattern.}
\Cref{fig:intra-session-arrival-cdf} first shows the CDF of intervals between
consecutive requests within the same session.  Most turns arrive in quick
succession: the median interval is only $20.8\,$s and $90\%$ of intervals are
below $120.6\,$s.  These short inter-arrival times help explain the high KV
cache hit rates above, since a session's recently produced KV cache is likely
to remain available when its next turn arrives.  The distribution nevertheless
has a long tail: the $99^{\mathrm{th}}$ percentile reaches $594.1\,$s, owing to
agent actions or user interaction.  A hierarchical memory system can retain
such inactive session KV caches in higher-capacity tiers, reducing
capacity-driven evictions as the number of sessions increases while keeping their
prefixes available for later reuse.

\Cref{fig:request-volume-hitrate-trend} further shows pronounced diurnal
variation in request volume, with a low-load period followed by sharp ramp-ups
and a sustained high-load window.  During some rapid increases in request
volume (e.g., from 06:30 to 08:00), the KV cache hit rate drops as new or cold
sessions comprise a larger share of arriving requests.  Such requests cannot
benefit from high KV cache reuse and must perform more prefill computation,
inflating per-request service demand and thus lowering the effective service
rate $\mu$ exactly when the arrival rate $\lambda$ is rising.  By
\Cref{eq:rho}, utilization climbs on both counts, and the resulting queueing
pressure makes these transition periods particularly important for scheduling.

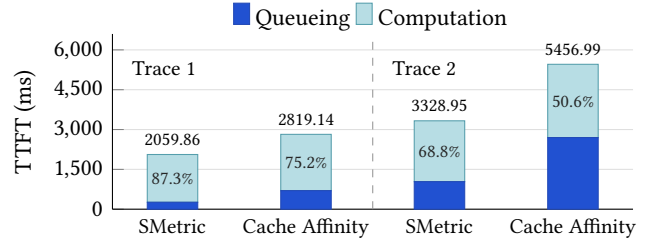
\begin{figure}[!tb]
  \centering
  \begin{tikzpicture}
    \begin{axis}[
        width=\columnwidth,
        height=3.9cm,
        ybar stacked,
        bar width=19pt,
        every axis plot/.append style={bar shift=0pt},
        ymin=0, ymax=6600,
        ytick={0,1500,3000,4500,6000},
        ylabel={TTFT (ms)},
        ylabel style={font=\small},
        symbolic x coords={S1,A1,S2,A2},
        xtick=data,
        xticklabels={SMetric, Cache Affinity, SMetric, Cache Affinity},
        xticklabel style={font=\footnotesize, align=center},
        yticklabel style={font=\small},
        axis lines*=left,
        ymajorgrids=true,
        grid style={draw=vestaGray!30},
        enlarge x limits=0.15,
        clip=false,
        legend columns=-1,
        legend style={font=\small, at={(0.5,0.98)}, anchor=south,
                      draw=none, fill=none, inner sep=1pt},
      ]
      \addplot[draw=vestaBlue!85!vestaDark, fill=vestaBlue]
        table[x=cfg, y=queue] {
        cfg   queue    compute  total
        S1    261.52   1798.34  2059.86
        A1    699.76   2119.38  2819.14
        S2    1038.00  2290.95  3328.95
        A2    2697.80  2759.19  5456.99
      };
      \addlegendentry{Queueing}
      \addplot[draw=vestaCyan!85!vestaDark, fill=vestaCyan!45]
        table[x=cfg, y=compute] {
        cfg   queue    compute  total
        S1    261.52   1798.34  2059.86
        A1    699.76   2119.38  2819.14
        S2    1038.00  2290.95  3328.95
        A2    2697.80  2759.19  5456.99
      };
      \addlegendentry{Computation}
      \draw[dashed, draw=vestaGray]
        (rel axis cs:0.5,0) -- (rel axis cs:0.5,1);
      \coordinate (ttftS1top) at (axis cs:S1,2059.86);
      \coordinate (ttftA1top) at (axis cs:A1,2819.14);
      \coordinate (ttftS2top) at (axis cs:S2,3328.95);
      \coordinate (ttftA2top) at (axis cs:A2,5456.99);
      \coordinate (ttftS1share) at (axis cs:S1,1160.69);
      \coordinate (ttftA1share) at (axis cs:A1,1759.45);
      \coordinate (ttftS2share) at (axis cs:S2,2183.475);
      \coordinate (ttftA2share) at (axis cs:A2,4077.395);
      \coordinate (ttftTraceOne) at (rel axis cs:0.02,0.90);
      \coordinate (ttftTraceTwo) at (rel axis cs:0.52,0.90);
    \end{axis}
    \node[font=\scriptsize, anchor=south] at (ttftS1top) {2059.86};
    \node[font=\scriptsize, anchor=south] at (ttftA1top) {2819.14};
    \node[font=\scriptsize, anchor=south] at (ttftS2top) {3328.95};
    \node[font=\scriptsize, anchor=south] at (ttftA2top) {5456.99};
    \node[font=\scriptsize, text=vestaDark] at (ttftS1share) {87.3\%};
    \node[font=\scriptsize, text=vestaDark] at (ttftA1share) {75.2\%};
    \node[font=\scriptsize, text=vestaDark] at (ttftS2share) {68.8\%};
    \node[font=\scriptsize, text=vestaDark] at (ttftA2share) {50.6\%};
    \node[font=\footnotesize, anchor=north west] at (ttftTraceOne) {Trace 1};
    \node[font=\footnotesize, anchor=north west] at (ttftTraceTwo) {Trace 2};
  \end{tikzpicture}
  \caption{TTFT decomposition on two traces. Queueing delay is represented by the dark segment, and the computation share is shown inside the light segment.}
  \label{fig:ttft-queue}
\end{figure}

\subsection{Queueing Effects of Cache-Aware Scheduling}

To quantify how cache-aware routing affects queueing, we replay two one-hour
traces---\emph{Trace 1} with 2,779 requests and \emph{Trace 2} with 4,036
requests---on Qwen2.5-7B-Instruct-1M~\cite{qwen2.5-1m}, based on their
original request timestamps.  The two traces represent different load regimes,
allowing us to examine queueing effects under different levels of load pressure.
The serving cluster comprises four prefill and four decode instances (4P4D),
each equipped with one H20 GPU and 128~GB of host DRAM.  We compare two
scheduling policies.  \emph{Cache Affinity} pins all requests in a session to
the instance that serves its first request, maximizing local KV cache reuse.
\emph{SMetric}~\cite{arxiv26_smetric} is a session-centric policy that
load-balances a session's first request and routes follow-up requests to the
same instance for cache affinity; it migrates a session's prefill to another
instance only after its cache-affine instance becomes overloaded.

\Cref{fig:ttft-queue} decomposes TTFT under the two policies into queueing and
computation time.  At the lower average cluster utilization of $29\%$ in
\emph{Trace 1}, SMetric reduces average TTFT from 2,819.14 to 2,059.86~ms
($26.9\%$).  More importantly, queueing falls from 699.76~ms ($24.8\%$ of
TTFT) under Cache Affinity to 261.52~ms ($12.7\%$) under SMetric.  At the
higher $45\%$ utilization of \emph{Trace 2}, this gap widens: SMetric reduces
TTFT from 5,456.99 to 3,328.95~ms ($39.0\%$) and queueing from 2,697.80~ms
($49.4\%$ of TTFT) to 1,038.00~ms ($31.2\%$).  Hence, as utilization rises,
queueing becomes the dominant source of Cache Affinity's TTFT inflation.  
Although SMetric alleviates this effect, its reactive, overload-triggered migration leaves substantial queueing delay,
particularly in \emph{Trace 2}.

To understand the remaining opportunity, consider $c$ homogeneous prefill instances with
aggregate arrival rate $\lambda$ and per-instance service rate $\mu$.  Strict
Cache Affinity partitions the cluster into $c$ logical queues: each session is
pinned to one instance, so the system is approximated by $c$ independent
\mbox{M/M/1} queues.  Under balanced average assignment, each queue receives
$\lambda/c$ requests per unit time and has utilization
\begin{equation}
\label{eq:affinity-rho}
  \rho = \frac{\lambda}{c\mu} < 1,
\end{equation}
which yields the mean queueing delay
\begin{equation}
\label{eq:wq-affinity}
  W_q^{\mathrm{aff}} = \frac{\rho}{\mu(1-\rho)}.
\end{equation}
Although the average assignment is balanced, the per-session pinning prevents
an idle instance from serving work queued at a busy cache holder.  The cluster
therefore forgoes the pooling gain that a shared queue would provide.

The fully pooled \mbox{M/M/$c$} queue provides an idealized reference point for
the opportunity missed by SMetric.  Let $a = \lambda/\mu = c\rho$ denote the offered load.  In this
model, the Erlang-C formula~\cite{queueing_theory} gives the probability that an arriving request must
wait:
\begin{equation}
\label{eq:erlang-c}
  P_{\mathrm{wait}} =
  \frac{\dfrac{a^c}{c!}\dfrac{1}{1-\rho}}
  {\displaystyle\sum_{n=0}^{c-1}\frac{a^n}{n!}
   + \dfrac{a^c}{c!}\dfrac{1}{1-\rho}}.
\end{equation}
Its mean queueing delay is therefore
\begin{equation}
\label{eq:wq-mmc}
  W_q^{\mathrm{pool}} = \frac{P_{\mathrm{wait}}}{c\mu-\lambda}
                        = \frac{P_{\mathrm{wait}}}{c\mu(1-\rho)},
\end{equation}
and the resulting relative delay is
\begin{equation}
\label{eq:pooling-gain}
  \frac{W_q^{\mathrm{pool}}}{W_q^{\mathrm{aff}}} =
  \frac{P_{\mathrm{wait}}}{c\rho}.
\end{equation}
For four homogeneous prefill instances at the same average per-instance
utilization, the ideal pooled \mbox{M/M/4} model predicts mean queueing delays
only $2.9\%$ and $7.1\%$ of those under Cache Affinity at $29\%$ and $45\%$
utilization, respectively.  SMetric's overload-triggered migration can capture
only part of this benefit, because it retains session pinning until a queue has
already become overloaded.

\section{Vesta: Queueing-Aware Scheduling}
\label{sec:schedule}

\begin{figure}[!tb]
  \centering
  \begin{subfigure}[b]{0.44\columnwidth}
    \centering
    \includegraphics[width=\linewidth]{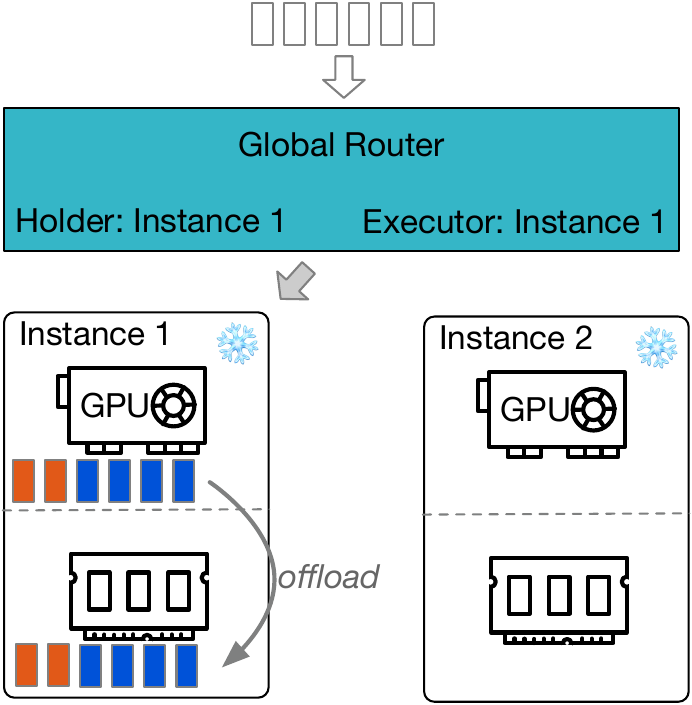}
    \caption{}
    \label{fig:holder-executor1}
  \end{subfigure}\hfill
  \begin{subfigure}[b]{0.44\columnwidth}
    \centering
    \includegraphics[width=\linewidth]{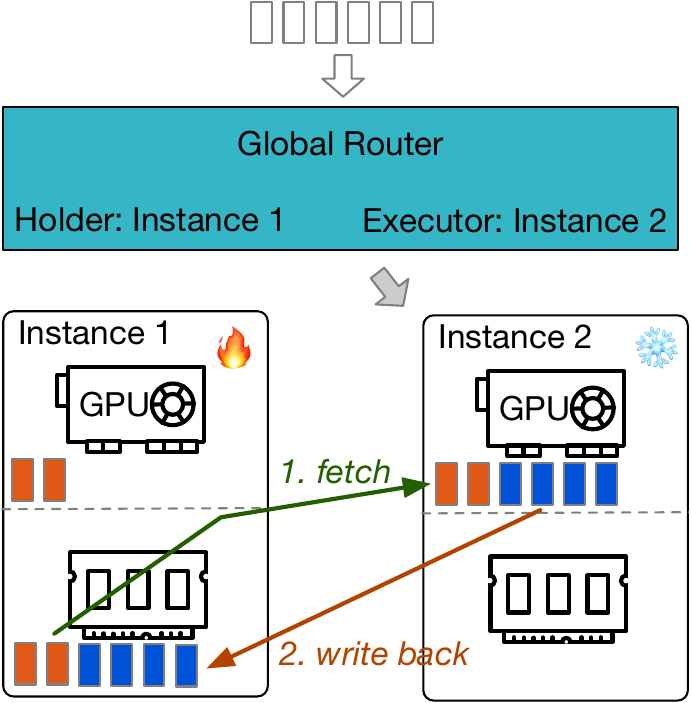}
    \caption{}
    \label{fig:holder-executor2}
  \end{subfigure}
  \caption{Illustration of the Holder/Executor-disaggregated architecture. (a) The holder and executor are assigned to the same instance. (b) The holder and executor are assigned to different instances. The \textcolor{vestaOrange}{orange blocks} represent the matched prefix cache, while the \textcolor{vestaBlue}{blue blocks} indicate the newly generated KV cache.}
  \label{fig:holder-executor}
\end{figure}

\subsection{Holder-Executor Disaggregation}
\label{sec:schedule:architecture}

Before presenting the scheduling algorithm, we describe Vesta's serving
architecture.  Our workload characterization exposes a fundamental tension
between cache affinity and load balance.  Strong intra-session KV cache reuse
eliminates most prefill computation for follow-up requests, reducing server
utilization and increasing the effective service rate.  Cache-affine routing
exploits this benefit by directing a session's requests to the instance that
holds its KV cache.  However, it can create severe local queueing when that
instance is busy, even if other instances have spare capacity.  The global
L2 cache hierarchy provides a way to resolve this tension: another instance can
fetch and reuse the prefix remotely, paying a transfer cost rather than
recomputing it.

Vesta exploits this capability by decoupling \emph{KV cache ownership} from
\emph{request execution}.  As shown in \Cref{fig:holder-executor}, each session
$s$ is associated with a \emph{holder} $h(s)$, whose role is to maintain the
authoritative KV cache of $s$ across turns.  For each request, Vesta selects an
\emph{executor} $e$ from the serving instances to perform the computation.  The
executor can be the holder itself: when $e=h(s)$
(\Cref{fig:holder-executor1}), it reads the reusable KV cache locally and
executes without a network transfer.  Alternatively, when the holder is busy
and another instance offers lower queueing pressure, Vesta can choose
$e\neq h(s)$ (\Cref{fig:holder-executor2}).  The remote executor fetches the
reusable prefix from the holder's L2 cache into its HBM, performs the prefill
computation, and writes the newly generated suffix KV cache back to the holder.

Notably, the remote executor does not offload either the fetched prefix KV
cache or the newly generated suffix KV cache to its local L2 cache. 
Such replicas would occupy L2 cache that could instead hold the KV cache of the sessions this instance holds, thereby reducing the cluster's effective cache capacity.
Instead, these KV caches remain in HBM and can be reused if a subsequent
request from the same session is again assigned to this executor.  The executor
then fetches only the newly missing suffix from the holder, substantially
reducing the remote-transfer cost of subsequent execution migration when the holder is
busy and this remote executor remains lightly loaded.
Because L2 cache capacity is much larger than HBM capacity, retaining
execution-side copies in HBM does not reduce the L2 capacity available for the
instance to serve as a holder for other sessions.

This write-back protocol preserves the holder as the owner of the
longest KV cache accumulated for its session despite dynamic execution
placement.  Subsequent turns can therefore always locate the session's complete
KV cache at $h(s)$, while Vesta remains free to migrate the current turn to a
less-loaded executor.  Holder-executor disaggregation thus retains the
long-term cache locality required by agentic sessions while recovering the
short-term capacity-pooling opportunity needed to reduce queueing.  The
scheduler must then decide when the queueing reduction from execution migration exceeds
its remote KV-cache transfer cost; we describe this decision next.

\subsection{Session-Balanced Holder Assignment}
\label{sec:schedule:holder-assignment}

Holder-executor disaggregation separates placement into two timescales: a
holder binding that persists for the lifetime of a session, and an executor
choice made anew for each of its requests.
Because the executor can adapt to instantaneous queueing conditions without
changing KV cache ownership, holder assignment can instead prioritize the
long-term distribution of sessions.  Vesta assigns each session to a holder by
hashing its session ID.  Specifically, given an ordered holder set
$\mathcal{H}=(h_0,\ldots,h_{c-1})$, Vesta defines
\begin{equation}
\label{eq:holder-assignment}
  h(s) = h_{\mathrm{Hash}(s) \bmod c},
\end{equation}
where $\mathrm{Hash}$ is a uniform hash function and $c=|\mathcal{H}|$.  This
mapping makes the holder of a session deterministic and independent of request
arrival order or transient load conditions.

To characterize its long-term balancing property, consider $N$ sessions assigned
over a sufficiently long period and let $c = |\mathcal{H}|$ be the number of
holders.  Under a uniform hash, the number of sessions assigned to any holder
$i$, denoted by $X_i$, follows
\begin{equation}
\label{eq:holder-session-count}
  X_i \sim \mathrm{Binomial}\!\left(N, \frac{1}{c}\right), \qquad
  \mathbb{E}[X_i] = \frac{N}{c}.
\end{equation}
Moreover, for any $0 < \epsilon \leq 1$, a Chernoff bound~\cite{mitzenmacher2005probability} gives
\begin{equation}
\label{eq:holder-balance-bound}
  \Pr\!\left[\left|X_i - \frac{N}{c}\right|
  \geq \epsilon\frac{N}{c}\right]
  \leq 2\exp\!\left(-\frac{\epsilon^2 N}{3c}\right).
\end{equation}
Thus, as the number of assigned sessions grows, each holder's session count
concentrates around $N/c$.  This long-term guarantee concerns session count,
not instantaneous request load, which is handled separately by executor
selection.

Hash-based assignment also eliminates the need for a centralized
session-to-holder mapping: given a session ID and a consistent ordered holder
set, any scheduler can independently compute $h(s)$.  This stateless property
simplifies scheduler recovery and supports decentralized routing, facilitating
large-scale deployment.

\paragraph{Cross-Bucket KV Cache Migration.}
Agentic sessions lengthen monotonically across turns as model responses and tool
outputs are appended to the context.  Consequently, a session may move to a
larger size-homogeneous bucket over time, e.g., from a 64K-token bucket to a
256K-token bucket.  When a request is new to its target bucket, Vesta queries
the session's bucket-level metadata.  If the session has a holder in a smaller
bucket, Vesta identifies the holder with the longest available KV cache prefix
and uses it as the migration source.

Vesta also assigns the session a holder in the target bucket through session hashing.
To avoid cold-starting the target bucket, Vesta transfers the reusable prefix cache from the source holder to
the two-tier cache of the target-bucket holder.
The target holder then retains the
newly generated suffix KV cache for subsequent requests in that bucket.
In this way, it preserves the session cache accumulated before the bucket transition instead of recomputing the entire context.

\begin{algorithm}[!tb]
  \caption{Queueing-aware executor selection}
  \label{alg:executor-selection}
  \begin{algorithmic}[1]
    \Require Request $r$ of session $s$ with $N$ prompt tokens, holder set
    $\mathcal{H}$, executor set $\mathcal{E}$
    \State $h(s)\gets h_{\mathrm{Hash}(s)\bmod|\mathcal{H}|}$
      \Comment{assign on the first turn; recompute thereafter}
    \State $(H,\{G_i,B_i\}_{i\in\mathcal{E}},\rho)\gets
      \textsc{CollectState}(r)$
    \ForAll{$i\in\mathcal{E}$}
      \State $R_i\gets\max\{G_i,H\}$; \quad
      $D_i\gets R_i-G_i$
      \State $S_i\gets T_{\mathrm{prefill}}(R_i,N-R_i)+\alpha_i D_i$
    \EndFor
    \State $S_{\min}\gets\min_{i\in\mathcal{E}}S_i$; \quad
      $\pi\gets\textsc{CongestionPrice}(\rho)$
    \State $e\gets\arg\min_{i\in\mathcal{E}}
      \{B_i+S_i+\pi(S_i-S_{\min})\}$
    \State \Return $(h(s),e)$ \Comment{write the suffix KV cache back to $h(s)$}
  \end{algorithmic}
\end{algorithm}

\subsection{Queueing-Aware Executor Selection}
\label{sec:schedule:executor-selection}

After fixing a session's holder, Vesta selects an executor for each request by
jointly considering its predicted completion time and the additional service
demand introduced by its placement.  This distinction is essential.  Routing a
request away from a busy holder can reduce its immediate queueing delay, but it
may also require extra KV-cache movement and prefill work.  The latter increases
cluster utilization and, as \Cref{eq:wq-mm1,eq:wq-mmc} show, can sharply
increase queueing delay as the system approaches saturation.

For a request with $N$ prompt tokens, let $G_i$ denote the longest reusable
prefix already resident in candidate executor $i$'s HBM, and let $H$ denote the
longest reusable prefix available anywhere in the cluster's L2 cache, located
through the KV cache indexer.  Once the holder has served a turn of the
session, the write-back protocol makes its copy the longest, so $H$ resides at
$h(s)$.  Two cases depart from this.  On a session's first turn the holder has
cached nothing for it, so $H$ is at best a prefix shared across sessions, such
as a common system prompt, which some other instance may hold; the executor's
own HBM may likewise already hold such a prefix, making $G_i$ the larger of the
two.  The holder may also have evicted part of the session's KV cache.
Vesta therefore takes the longer of the two as the prefix that executor $i$
reuses:
\begin{equation}
\label{eq:executor-reuse}
  R_i = \max\{G_i, H\}, \qquad
  D_i = R_i-G_i,
\end{equation}
where $D_i\ge 0$ is the number of KV-cache tokens that must be loaded into
$i$'s HBM.  Pulling from an instance other than $h(s)$ does not affect
ownership: the executor writes back whatever portion of the context the holder
does not already hold, so $h(s)$ again covers the session in full and remains
its single accumulation point.  Vesta estimates the service demand of placing the request at $i$ as
\begin{equation}
\label{eq:executor-service-cost}
  S_i = T_{\mathrm{prefill}}(R_i, N-R_i) + \alpha_i D_i.
\end{equation}
Here, $T_{\mathrm{prefill}}(R_i,N-R_i)$ is an offline-calibrated prefill-latency
model for a request with $R_i$ cached and $N-R_i$ uncached tokens, and
$\alpha_i$ is the per-token cost of loading KV cache into $i$'s HBM, which is
lower when $i$ reads from its own host memory than when it fetches from another
instance over the network.
Thus, $S_i$ captures both the computation
saved by KV cache reuse and the transfer work required to make a reusable prefix
available at the executor.

Let $B_i$ be the predicted unfinished prefill work at executor $i$.  Let
$S_{\min}$ denote the minimum service cost across candidates.  The end-to-end
procedure is summarized in \Cref{alg:executor-selection}.  Vesta selects the
executor that minimizes a congestion-priced completion-time objective:
\begin{equation}
\label{eq:executor-objective}
  \begin{aligned}
    e^* = \arg\min_{i \in \mathcal{E}} \bigl\{&B_i + S_i \\
      &+ \pi(\rho)\bigl(S_i-S_{\min}\bigr)\bigr\}.
  \end{aligned}
\end{equation}
Here, $B_i+S_i$ estimates the request's completion time at $i$.  Minimizing
this term alone is insufficient: it optimizes the current request but ignores
the queueing externality of placing extra service demand on a shared cluster.
In an \mbox{M/M/1} queue, for example, the sensitivity of mean queueing delay to
utilization is
\begin{equation}
\label{eq:queueing-sensitivity}
  \frac{\partial W_q}{\partial\rho}
  = \frac{1}{\mu(1-\rho)^2},
\end{equation}
which diverges as $\rho\to1$.  The same convexity arises in \mbox{M/M/$c$}: an
incremental increase in service demand has little system-wide effect at low
utilization, but delays many subsequent requests when the cluster is congested.
The final term in \Cref{eq:executor-objective} prices this externality.  It
penalizes only $S_i-S_{\min}$, i.e., the service demand that placing the
request at $i$ adds over the least-demanding candidate, while leaving the
unavoidable cost $S_{\min}$ unpriced.

Vesta uses a bounded, queueing-theory-inspired surrogate for this price.  It is
derived from the cluster-level forward-time utilization $\rho$ and the
\mbox{M/M/$c$} waiting probability in \Cref{eq:erlang-c}:
\begin{equation}
\label{eq:congestion-price}
  \pi(\rho) = \min\!\left\{\pi_{\max},
    \frac{\rho}{1-\rho}
    \left(P_{\mathrm{wait}}^{(c)}(\rho)\right)^\gamma\right\}.
\end{equation}
The factor $\rho/(1-\rho)$ captures the rapidly increasing marginal queueing
cost near saturation, while $P_{\mathrm{wait}}^{(c)}(\rho)$ suppresses the
price when an arriving request is unlikely to wait in the pooled cluster.
$\gamma$ controls how sharply this waiting probability affects the price, and
$\pi_{\max}$ bounds the influence of measurement and modeling error.  Thus, at
low cluster utilization, $\pi(\rho)$ is near zero and Vesta can migrate requests
aggressively to recover idle capacity from other executors.  As utilization and
queue pressure rise, the price increasingly favors placements with smaller
$S_i$, such as the holder itself or other instances with relatively long KV
cache prefixes. 
By avoiding additional KV-cache transfer work, these placements limit the
incremental service demand and help prevent further increases in $\rho$.

Vesta more closely approaches the capacity pooling of an \mbox{M/M/$c$} queue
than SMetric.  SMetric preserves cache affinity until the holder becomes
overloaded, so requests can remain queued at a busy holder while other executors
have capacity.  In contrast, Vesta evaluates every executor for every request:
the backlog term $B_i$ exposes idle capacity beyond the holder, while the
congestion-priced service term triggers execution migration only when its
queueing benefit justifies the additional KV-cache transfer demand.  Vesta can therefore exploit
capacity-pooling opportunities before a holder becomes overloaded without
blindly increasing $\rho$, bringing its queueing behavior closer to the pooled
\mbox{M/M/$c$} model.

\section{Evaluation}
\label{sec:eval}

\subsection{Experimental Setup}
\noindent\textbf{Implementation.}
Our cluster comprises 12 GPU servers totally, each equipped with 8 NVIDIA H20 GPUs (96~GB HBM each), 96 CPU cores, and 1,024~GB of DRAM.
The servers are interconnected via 200~Gb/s RDMA NICs.
The instances in the cluster are running the same vLLM engine~\cite{vllm}.
Dynamo~\cite{ai-dynamo} serves as the KV cache indexer, tracking the locations
of KV cache blocks across instances to support cache lookup and prefill latency
estimation. Empirically, we set the congestion-price exponent $\gamma=0.5$ and cap
$\pi_{\max}=2$ for Vesta in all experiments.

\noindent\textbf{Models.}
We evaluate Qwen2.5-7B-Instruct-1M~\cite{qwen2.5-1m} and
Hy3-FP8~\cite{hy3}, two models at different scales that both support long
contexts.
Each Qwen2.5-7B-Instruct-1M instance uses one GPU, whereas each Hy3-FP8 instance uses eight GPUs.

\noindent\textbf{Real-World Agentic Workloads.}
We collect \emph{Trace Agent}, a one-hour trace of peak-time real-world agentic workloads (36K requests in total, mean input length: 83K, P90 input length: 133K), from an online serving system.
In all experiments, we replay the trace
according to its original timestamps to preserve the observed request-arrival
pattern, while serializing requests within each session to prevent concurrent
processing of dependent turns.
Following prior work~\cite{osdi26_lmetric,arxiv26_smetric}, we force each
evaluated model to generate exactly the number of tokens recorded in the trace.

\noindent\textbf{Baselines.}
We compare Vesta with the state-of-the-art LLM scheduling strategies including LMetric~\cite{osdi26_lmetric} and SMetric~\cite{arxiv26_smetric}.
LMetric uses the product of the number of uncached tokens and batch size as its
scheduling score.  SMetric adopts a session-centric policy that load-balances
first-turn requests and routes follow-up requests to the same instance for cache
affinity, unless that instance is overloaded or the session's KV cache has been
evicted.

\subsection{Serving Performance in PD-Colocated Architecture}
\label{sec:eval-pd-colocated}

We evaluate Qwen2.5-7B-Instruct-1M with 32 instances on \emph{Trace Agent}, each colocating
prefill and decode execution.

\begin{figure}[!tb]
  \centering
  \captionsetup[subfigure]{font=scriptsize,skip=2pt}
  \pgfplotsset{
    ifb axis/.style={
      scale only axis,
      width=0.70\linewidth,
      height=2.1cm,
      ymin=0,
      axis lines*=left,
      ymajorgrids=true,
      grid style={draw=vestaGray!30},
      tick align=outside,
      tick label style={font=\fontsize{6}{7}\selectfont},
      label style={font=\scriptsize},
      scaled y ticks=true,
      y tick scale label style={font=\fontsize{6}{7}\selectfont},
      yticklabel style={/pgf/number format/fixed},
      clip=false,
    },
    ifb lmetric/.style={
      draw=vestaOrange!85!vestaDark,
      fill=vestaOrange!30,
      postaction={pattern=north east lines, pattern color=vestaOrange!85!vestaDark},
    },
    ifb smetric/.style={
      draw=vestaCyan!80!vestaDark,
      fill=vestaCyan!30,
      postaction={pattern=horizontal lines, pattern color=vestaCyan!80!vestaDark},
    },
    ifb vesta/.style={draw=vestaBlue!85!vestaDark, fill=vestaBlue!85},
    ifb values/.style={
      nodes near coords,
      every node near coord/.append style={
        font=\fontsize{5.5}{6.5}\selectfont,
        text=vestaDark,
        /pgf/number format/fixed,
        /pgf/number format/precision=1,
        /pgf/number format/fixed zerofill,
      },
    },
    ifb tail/.style={
      ifb axis,
      ybar=0.5pt,
      bar width=3.5pt,
      symbolic x coords={P50,P90,P95,P99},
      xtick=data,
      enlarge x limits=0.19,
    },
    ifb mean/.style={
      ifb axis,
      ybar=0pt,
      bar width=6pt,
      symbolic x coords={TTFT,Queueing},
      xtick=data,
      enlarge x limits=0.6,
    },
  }
  \pgfplotslegendfromname{sota-ifb-legend}\par\smallskip
  \begin{subfigure}[t]{0.48\columnwidth}
    \centering
    \begin{tikzpicture}
      \begin{axis}[
          ifb mean,
          ymax=6,
          ytick={0,2,4,6},
          ylabel={Mean latency (s)},
          legend to name=sota-ifb-legend,
          legend columns=3,
          legend style={font=\scriptsize, draw=none, fill=none,
                        /tikz/every even column/.append style={column sep=6pt}},
        ]
        \addplot+[ifb lmetric] coordinates {(TTFT,5.56887) (Queueing,0.29604)};
        \addlegendentry{LMetric}
        \addplot+[ifb smetric] coordinates {(TTFT,3.99389) (Queueing,1.46364)};
        \addlegendentry{SMetric}
        \addplot+[ifb vesta] coordinates {(TTFT,1.68563) (Queueing,0.06391)};
        \addlegendentry{Vesta}
      \end{axis}
    \end{tikzpicture}
    \caption{Mean TTFT and queueing delay.}
    \label{fig:sota-ifb-mean}
  \end{subfigure}\hfill
  \begin{subfigure}[t]{0.48\columnwidth}
    \centering
    \begin{tikzpicture}
      \begin{axis}[
          ifb axis,
          ifb values,
          ybar,
          bar width=12pt,
          bar shift=0pt,
          xmin=0.4, xmax=3.6,
          xtick={1,2,3},
          xticklabels={LMetric,SMetric,Vesta},
          ymax=60,
          ytick={0,15,30,45,60},
          ylabel={TPS (tokens/s)},
        ]
        \addplot+[ifb lmetric] coordinates {(1,38.155)};
        \addplot+[ifb smetric] coordinates {(2,46.873)};
        \addplot+[ifb vesta] coordinates {(3,53.417)};
      \end{axis}
    \end{tikzpicture}
    \caption{Tokens per second (TPS).}
    \label{fig:sota-ifb-tps}
  \end{subfigure}\par\medskip
  \begin{subfigure}[t]{0.48\columnwidth}
    \centering
    \begin{tikzpicture}
      \begin{axis}[
          ifb tail,
          ymax=45,
          ytick={0,15,30,45},
          ylabel={TTFT (s)},
        ]
        \addplot+[ifb lmetric] coordinates {
          (P50,1.790) (P90,16.171) (P95,20.337) (P99,32.7138)};
        \addplot+[ifb smetric] coordinates {
          (P50,0.645) (P90,12.5494) (P95,19.6126) (P99,40.36672)};
        \addplot+[ifb vesta] coordinates {
          (P50,0.393) (P90,3.9482) (P95,8.0696) (P99,20.45744)};
      \end{axis}
    \end{tikzpicture}
    \caption{TTFT percentiles.}
    \label{fig:sota-ifb-ttft}
  \end{subfigure}\hfill
  \begin{subfigure}[t]{0.48\columnwidth}
    \centering
    \begin{tikzpicture}
      \begin{axis}[
          ifb tail,
          ymax=30,
          ytick={0,10,20,30},
          ylabel={Queueing delay (s)},
        ]
        \addplot+[ifb lmetric] coordinates {
          (P50,0) (P90,0) (P95,1.077) (P99,8.065)};
        \addplot+[ifb smetric] coordinates {
          (P50,0) (P90,3.5044) (P95,10.380) (P99,28.14352)};
        \addplot+[ifb vesta] coordinates {
          (P50,0) (P90,0) (P95,0) (P99,0)};
      \end{axis}
    \end{tikzpicture}
    \caption{Queueing-delay percentiles.}
    \label{fig:sota-ifb-queue}
  \end{subfigure}
  \caption{Serving performance on Qwen2.5-7B-Instruct-1M with 32
  PD-colocated instances.}
  \Description{Four bar charts compare LMetric, SMetric, and Vesta on
  Qwen2.5-7B-Instruct-1M with 32 PD-colocated vLLM instances.
  Vesta has the lowest mean TTFT and queueing delay,
  the lowest TTFT at all four reported percentiles, and the highest TPS:
  53.4 tokens per second versus 46.9 for SMetric and 38.2 for LMetric.
  Vesta's queueing percentiles from P50 through P99 are zero, although its
  mean queueing delay is nonzero at 0.06 seconds.}
  \label{fig:sota-compare-ifb}
\end{figure}
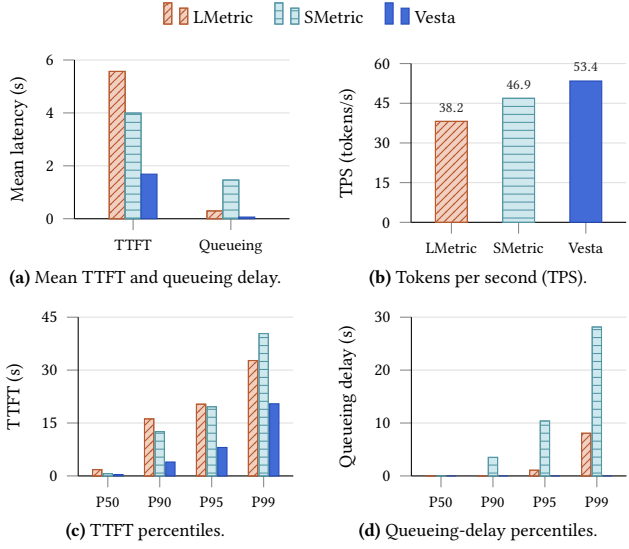

\Cref{fig:sota-compare-ifb} compares the serving performance of LMetric,
SMetric, and Vesta in the PD-colocated setting. Vesta achieves both the
highest tokens per second (TPS) and the lowest mean TTFT. \Cref{fig:sota-ifb-tps} shows that
Vesta reaches 53.4~tokens/s, $13.9\%$ higher than SMetric
(46.9~tokens/s), the stronger baseline for TPS, and $39.8\%$ higher
than LMetric (38.2~tokens/s). This improvement does not come at the cost
of a longer wait for the first token: \Cref{fig:sota-ifb-mean} shows that
Vesta reduces mean TTFT to 1,685.63~ms, compared with 3,993.89~ms under
SMetric ($57.8\%$ lower) and 5,568.87~ms under LMetric ($69.7\%$ lower).

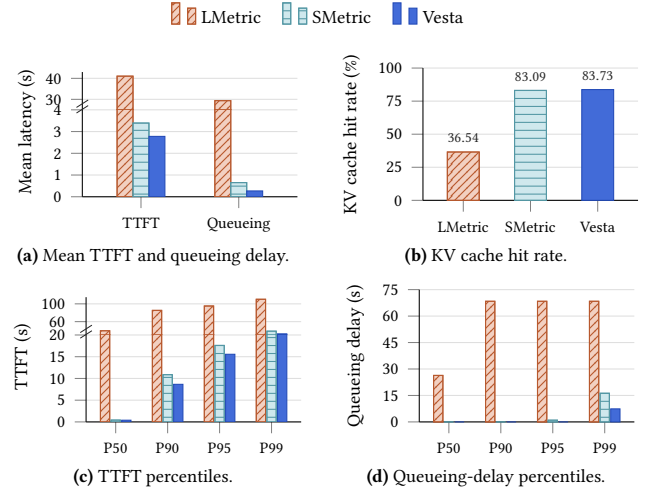
\begin{figure}[!tb]
  \centering
  \captionsetup[subfigure]{font=scriptsize,skip=2pt}
  \pgfplotsset{
    sota axis/.style={
      scale only axis,
      width=0.70\linewidth,
      height=1.75cm,
      ymin=0,
      axis lines*=left,
      ymajorgrids=true,
      grid style={draw=vestaGray!30},
      tick align=outside,
      tick label style={font=\fontsize{6}{7}\selectfont},
      label style={font=\scriptsize},
      scaled y ticks=true,
      y tick scale label style={font=\fontsize{6}{7}\selectfont},
      yticklabel style={/pgf/number format/fixed},
      clip=false,
    },
    sota lmetric/.style={
      draw=vestaOrange!85!vestaDark,
      fill=vestaOrange!30,
      postaction={pattern=north east lines, pattern color=vestaOrange!85!vestaDark},
    },
    sota smetric/.style={
      draw=vestaCyan!80!vestaDark,
      fill=vestaCyan!30,
      postaction={pattern=horizontal lines, pattern color=vestaCyan!80!vestaDark},
    },
    sota vesta/.style={draw=vestaBlue!85!vestaDark, fill=vestaBlue!85},
    sota values/.style={
      nodes near coords,
      every node near coord/.append style={
        font=\fontsize{5.5}{6.5}\selectfont,
        text=vestaDark,
        /pgf/number format/fixed,
        /pgf/number format/precision=2,
        /pgf/number format/fixed zerofill,
      },
    },
    sota tail/.style={
      sota axis,
      ybar=0.5pt,
      bar width=3.5pt,
      symbolic x coords={P50,P90,P95,P99},
      xtick=data,
      enlarge x limits=0.19,
    },
    sota mean/.style={
      sota axis,
      ybar=0pt,
      bar width=6pt,
      symbolic x coords={TTFT,Queueing},
      xtick=data,
      enlarge x limits=0.6,
    },
    sota break lower/.style={
      clip=true,
      scaled y ticks=false,
      height=1.15cm,
    },
    sota break upper/.style={
      clip=true,
      scaled y ticks=false,
      height=0.5cm,
      axis x line=none,
      anchor=south west,
    },
  }
  \providecommand{\sotabreakmark}[1]{%
    \draw[draw=vestaDark, line width=0.35pt]
      ([xshift=-2.2pt,yshift=-1.2pt]#1.north west) --
      ([xshift=2.2pt,yshift=1.6pt]#1.north west);
    \draw[draw=vestaDark, line width=0.35pt]
      ([xshift=-2.2pt,yshift=0.9pt]#1.north west) --
      ([xshift=2.2pt,yshift=3.7pt]#1.north west);
  }
  \pgfplotslegendfromname{sota-pd-legend}\par\smallskip
  \begin{subfigure}[t]{0.48\columnwidth}
    \centering
    \begin{tikzpicture}
      \begin{axis}[
          sota mean,
          sota break lower,
          name=sotameanlow,
          ymax=4,
          ytick={0,1,2,3,4},
          ylabel={Mean latency (s)},
          ylabel style={at={(ticklabel cs:0.75)}},
          legend to name=sota-pd-legend,
          legend columns=3,
          legend style={font=\scriptsize, draw=none, fill=none,
                        /tikz/every even column/.append style={column sep=6pt}},
        ]
        \addplot+[sota lmetric] coordinates {(TTFT,41.04053) (Queueing,29.36956)};
        \addlegendentry{LMetric}
        \addplot+[sota smetric] coordinates {(TTFT,3.39152) (Queueing,0.65120)};
        \addlegendentry{SMetric}
        \addplot+[sota vesta] coordinates {(TTFT,2.77805) (Queueing,0.26814)};
        \addlegendentry{Vesta}
      \end{axis}
      \begin{axis}[
          sota mean,
          sota break upper,
          at={([yshift=2.5pt]sotameanlow.north west)},
          ymin=25, ymax=43,
          ytick={30,40},
        ]
        \addplot+[sota lmetric] coordinates {(TTFT,41.04053) (Queueing,29.36956)};
        \addplot[draw=none, fill=none] coordinates {(TTFT,25) (Queueing,25)};
        \addplot[draw=none, fill=none] coordinates {(TTFT,25) (Queueing,25)};
      \end{axis}
      \sotabreakmark{sotameanlow}
    \end{tikzpicture}
    \caption{Mean TTFT and queueing delay.}
    \label{fig:sota-pd-mean}
  \end{subfigure}\hfill
  \begin{subfigure}[t]{0.48\columnwidth}
    \centering
    \begin{tikzpicture}
      \begin{axis}[
          sota axis,
          sota values,
          ybar,
          bar width=12pt,
          bar shift=0pt,
          xmin=0.4, xmax=3.6,
          xtick={1,2,3},
          xticklabels={LMetric,SMetric,Vesta},
          ymax=100,
          ytick={0,25,50,75,100},
          ylabel={KV cache hit rate (\%)},
        ]
        \addplot+[sota lmetric] coordinates {(1,36.54)};
        \addplot+[sota smetric] coordinates {(2,83.09)};
        \addplot+[sota vesta] coordinates {(3,83.73)};
      \end{axis}
    \end{tikzpicture}
    \caption{KV cache hit rate.}
    \label{fig:sota-pd-hit}
  \end{subfigure}\par\smallskip
  \begin{subfigure}[t]{0.48\columnwidth}
    \centering
    \begin{tikzpicture}
      \begin{axis}[
          sota tail,
          sota break lower,
          name=sotattftlow,
          ymax=20,
          ytick={0,5,10,15,20},
          ylabel={TTFT (s)},
          ylabel style={at={(ticklabel cs:0.75)}},
        ]
        \addplot+[sota lmetric] coordinates {
          (P50,39.5015) (P90,85.334) (P95,95.1891) (P99,110.3067)};
        \addplot+[sota smetric] coordinates {
          (P50,0.420) (P90,10.8724) (P95,17.59885) (P99,38.58485)};
        \addplot+[sota vesta] coordinates {
          (P50,0.394) (P90,8.6477) (P95,15.5618) (P99,32.577)};
      \end{axis}
      \begin{axis}[
          sota tail,
          sota break upper,
          at={([yshift=2.5pt]sotattftlow.north west)},
          ymin=30, ymax=115,
          ytick={60,100},
        ]
        \addplot+[sota lmetric] coordinates {
          (P50,39.5015) (P90,85.334) (P95,95.1891) (P99,110.3067)};
        \addplot+[sota smetric] coordinates {(P99,38.58485)};
        \addplot+[sota vesta] coordinates {(P99,32.577)};
      \end{axis}
      \sotabreakmark{sotattftlow}
    \end{tikzpicture}
    \caption{TTFT percentiles.}
    \label{fig:sota-pd-ttft}
  \end{subfigure}\hfill
  \begin{subfigure}[t]{0.48\columnwidth}
    \centering
    \begin{tikzpicture}
      \begin{axis}[
          sota tail,
          ymax=75,
          ytick={0,15,30,45,60,75},
          ylabel={Queueing delay (s)},
        ]
        \addplot+[sota lmetric] coordinates {
          (P50,26.361) (P90,68.3941) (P95,68.3941) (P99,68.3941)};
        \addplot+[sota smetric] coordinates {
          (P50,0) (P90,0) (P95,0.98055) (P99,16.25234)};
        \addplot+[sota vesta] coordinates {
          (P50,0) (P90,0) (P95,0) (P99,7.39711)};
      \end{axis}
    \end{tikzpicture}
    \caption{Queueing-delay percentiles.}
    \label{fig:sota-pd-queue}
  \end{subfigure}
  \caption{Serving performance on Qwen2.5-7B-Instruct-1M with 24P8D.}
  \Description{Four bar charts compare LMetric, SMetric, and Vesta on
  Qwen2.5-7B-Instruct-1M with 24 prefill and 8 decode instances.
  Vesta has the lowest mean and percentile TTFT and
  the lowest mean queueing delay, with queueing percentiles no higher than
  either baseline. Its cache hit rate is 83.73\%, compared with 83.09\%
   for SMetric and 36.54\% for LMetric. The y axes of the first and third
   charts are broken so that the much smaller SMetric and Vesta bars stay
   readable.}
  \label{fig:sota-compare-24p8d}
\end{figure}

\Cref{fig:sota-ifb-mean} also reports the queueing delay, which accounts for
much of this TTFT gap.
Although SMetric has lower mean TTFT than LMetric, its mean
queueing delay is higher: 1,463.64 versus 296.04~ms. Vesta lowers this delay
to 63.91~ms, a reduction of $95.6\%$ relative to SMetric and $78.4\%$
relative to LMetric. Consequently, the fraction of mean TTFT spent waiting
falls from $36.6\%$ under SMetric to $3.8\%$ under Vesta. 
It is consistent with Vesta's queueing-aware
executor selection: rather than preserving affinity until an overload
condition triggers migration, Vesta weighs each candidate's backlog against
its service cost for every request. Its congestion-priced objective
in~\Cref{eq:executor-objective} is designed to exploit capacity beyond a
busy holder while accounting for the additional service demand of migration.

The latency percentiles further distinguish the schedulers.
\Cref{fig:sota-ifb-ttft} shows that Vesta has the lowest TTFT at every
reported percentile. Relative to SMetric, Vesta reduces P90 and P95 TTFT
by $68.5\%$ and $58.9\%$, respectively. At P99, LMetric is the stronger
baseline, despite its higher mean TTFT: it reaches 32,713.80~ms versus
40,366.72~ms under SMetric. Vesta lowers P99 TTFT to 20,457.44~ms,
$37.5\%$ below LMetric and $49.3\%$ below SMetric.
\Cref{fig:sota-ifb-queue} shows zero queueing delay for Vesta at all four
reported percentiles, whereas P99 queueing reaches 8,065.00~ms under
LMetric and 28,143.52~ms under SMetric. These zero percentiles do not imply
that every request avoids queueing, as Vesta's mean queueing delay remains
nonzero. Together, the results show that Vesta improves TPS while reducing
both mean TTFT and the measured latency tail in this PD-colocated experiment.

Despite its higher TPS and lower mean TTFT, SMetric has $23.4\%$ higher
P99 TTFT and $3.49\times$ the P99 queueing delay of LMetric.
This reversal highlights that improvements in throughput and mean latency
do not necessarily translate into better tail performance.
This observation aligns with Vesta's motivation to jointly model KV cache transfer costs and their queueing externality: 
the additional service demand induced by migration can amplify waiting near saturation, even when migration benefits the current request.

\input{figs/load_line_chart}

\subsection{Serving Performance in PD-Disaggregated Architecture}
\label{sec:eval-pd-disaggregated}

\noindent\textbf{Qwen2.5-7B-Instruct-1M.}
We first evaluate Qwen2.5-7B-Instruct-1M on a 24P8D deployment with 2552 sessions (33K requests) sampled from \emph{Trace Agent}.

\Cref{fig:sota-compare-24p8d} compares the serving performance of LMetric,
SMetric, and Vesta in the 24P8D PD-disaggregated setting. Vesta achieves the
lowest mean TTFT and queueing delay among the three schedulers. Its mean
TTFT is 2,778.05~ms, $18.1\%$ lower than SMetric (3,391.52~ms), the
stronger baseline, and $93.2\%$ lower than LMetric (41,040.53~ms).

More importantly, Vesta lowers queueing delay while preserving KV cache
reuse. \Cref{fig:sota-pd-mean} shows that mean queueing delay falls from
651.20~ms under SMetric to 268.14~ms under Vesta ($58.8\%$), reducing its
share of mean TTFT from $19.2\%$ to $9.7\%$. Meanwhile,
\Cref{fig:sota-pd-hit} shows that their KV cache hit rates remain comparable:
$83.73\%$ for Vesta versus $83.09\%$ for SMetric. Relative to LMetric,
Vesta improves both metrics: its hit rate increases from $36.54\%$ to
$83.73\%$, while mean queueing delay decreases from 29,369.56 to
268.14~ms ($99.1\%$). Hence, the improvement over SMetric is accompanied
by substantially less waiting, rather than a large increase in cache hit rate.

These results are consistent with Vesta's holder--executor separation: it
uses capacity beyond a busy holder while preserving cache reuse through
write-back.  The benefit extends to the tail: relative to SMetric, Vesta
reduces P90, P95, and P99 TTFT by $20.5\%$, $11.6\%$, and $15.6\%$,
respectively, while lowering P99 queueing delay by $54.5\%$ (from 16,252.34
to 7,397.11~ms).

\begin{figure}[!tb]
  \centering
  \includegraphics[width=\columnwidth]{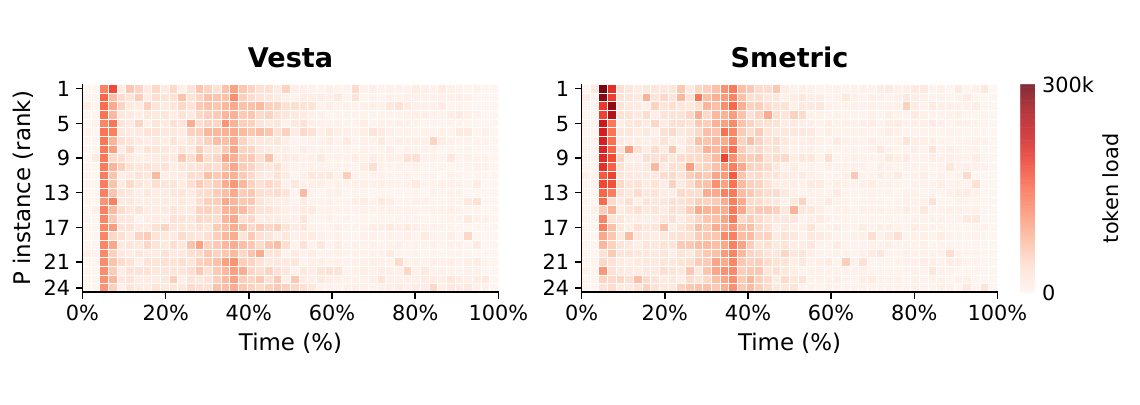}
  \caption{Prefill token load on Qwen2.5-7B-Instruct-1M (24P8D). Each cell
  reports the time-weighted mean inflight input tokens; instances are ranked by
  peak load.}
  \Description{Vesta (left) and SMetric (right) heatmaps for 24 prefill instances.
  SMetric has more intense hotspots on its highest-ranked instances;
  Vesta reduces the maximum instance-bin load from about 287 thousand
  to 176 thousand tokens.}
  \label{fig:prefill-load-qwen}
\end{figure}

\noindent\textbf{Prefill load balance.}
\Cref{fig:prefill-load-qwen} examines how the lower latency relates to
load distribution across prefill instances. We count a request's input
tokens from its logged \emph{P Process} event until completion or cancellation,
then take a time-weighted mean within each bin. Following the max-to-mean
comparison used by SMetric, we divide the maximum
instance load by the mean instance load in each bin and average these
ratios over fully observed bins with positive total load. A ratio closer
to one indicates more even load distribution.

Vesta weakens the load hotspots that SMetric exhibits during the high-load
period.
The maximum instance-bin load falls from 287.3k to 176.4k tokens
($38.6\%$), and the P99 load across observed instance-bin cells falls from
189.7k to 131.9k tokens ($30.5\%$). These are load statistics, not latency
percentiles. On the 44 corresponding bins with complete observations for
both policies, the mean max-to-mean ratio decreases from $3.045\times$ to
$2.987\times$ ($1.9\%$). Thus, the reduction in absolute hotspot intensity
is more pronounced than the improvement in relative balance. Together
with the comparable cache hit rates, these profiles are consistent with
Vesta using capacity beyond a busy holder without abandoning cache reuse.

\noindent\textbf{Performance under varying load.}
We evaluate the three schedulers on Qwen2.5-7B-Instruct-1M in the 24P8D
deployment at eight workload sizes from 15k to 36k requests from \emph{Trace Agent}.  \Cref{fig:load-line-chart}
reports mean, P90, and P99 TTFT and queueing delay; its queueing axes use
$\log_{10}(1+q)$ to retain zero-valued measurements.

Vesta achieves the lowest mean TTFT at every load.  Its advantage over SMetric
is modest at 15k--24k requests ($4.4\%$--$5.1\%$), but grows to $18.1\%$ and
$20.0\%$ at 33k and 36k, respectively.  From 30k to 36k, Vesta's mean TTFT
increases by 2,640.79~ms, $26.8\%$ less than SMetric's increase, while its mean
queueing delay increases by 753.64~ms, $47.7\%$ less.  At 33k and 36k, Vesta
also reduces mean queueing by $58.8\%$ and $45.8\%$ relative to SMetric.

The tail results show the same high-load benefit: at 33k and 36k, Vesta reduces
P90 TTFT by $20.5\%$ and $21.3\%$ and P99 TTFT by $15.6\%$ and $10.0\%$,
respectively.  Its P90 queueing remains zero through 36k, whereas SMetric
reaches 3,390.00~ms.  The advantage is not uniform at every intermediate load,
but Vesta consistently provides lower mean latency and substantially better
tail latency at the highest offered loads.
Furthermore, at a matched mean TTFT, Vesta sustains higher prefill throughput than the baselines.

\begin{figure}[!tb]
  \centering
  \captionsetup[subfigure]{font=scriptsize,skip=2pt}
  \pgfplotsset{
    hy3 axis/.style={
      scale only axis,
      width=0.70\linewidth,
      height=1.8cm,
      ymin=0,
      axis lines*=left,
      ymajorgrids=true,
      grid style={draw=vestaGray!30},
      tick align=outside,
      tick label style={font=\fontsize{6}{7}\selectfont},
      label style={font=\scriptsize},
      scaled y ticks=true,
      y tick scale label style={font=\fontsize{6}{7}\selectfont},
      yticklabel style={/pgf/number format/fixed},
      clip=false,
    },
    hy3 lmetric/.style={
      draw=vestaOrange!85!vestaDark,
      fill=vestaOrange!30,
      postaction={pattern=north east lines, pattern color=vestaOrange!85!vestaDark},
    },
    hy3 smetric/.style={
      draw=vestaCyan!80!vestaDark,
      fill=vestaCyan!30,
      postaction={pattern=horizontal lines, pattern color=vestaCyan!80!vestaDark},
    },
    hy3 vesta/.style={draw=vestaBlue!85!vestaDark, fill=vestaBlue!85},
    hy3 values/.style={
      nodes near coords,
      every node near coord/.append style={
        font=\fontsize{5.5}{6.5}\selectfont,
        text=vestaDark,
        /pgf/number format/fixed,
        /pgf/number format/precision=2,
        /pgf/number format/fixed zerofill,
      },
    },
    hy3 tail/.style={
      hy3 axis,
      ybar=0.5pt,
      bar width=3.5pt,
      symbolic x coords={P50,P90,P95,P99},
      xtick=data,
      enlarge x limits=0.19,
    },
    hy3 mean/.style={
      hy3 axis,
      ybar=0pt,
      bar width=6pt,
      symbolic x coords={TTFT,Queueing},
      xtick=data,
      enlarge x limits=0.6,
    },
    hy3 break lower/.style={
      clip=true,
      scaled y ticks=false,
      height=1.2cm,
    },
    hy3 break upper/.style={
      clip=true,
      scaled y ticks=false,
      height=0.5cm,
      axis x line=none,
      anchor=south west,
    },
  }
  \providecommand{\hybreakmark}[1]{%
    \draw[draw=vestaDark, line width=0.35pt]
      ([xshift=-2.2pt,yshift=-1.2pt]#1.north west) --
      ([xshift=2.2pt,yshift=1.6pt]#1.north west);
    \draw[draw=vestaDark, line width=0.35pt]
      ([xshift=-2.2pt,yshift=0.9pt]#1.north west) --
      ([xshift=2.2pt,yshift=3.7pt]#1.north west);
  }
  \pgfplotslegendfromname{sota-hy3-legend}\par\smallskip
  \begin{subfigure}[t]{0.48\columnwidth}
    \centering
    \begin{tikzpicture}
      \begin{axis}[
          hy3 mean,
          hy3 break lower,
          name=hy3meanlow,
          ymax=1.8,
          ytick={0,0.6,1.2},
          ylabel={Mean latency (s)},
          ylabel style={at={(ticklabel cs:0.75)}},
          legend to name=sota-hy3-legend,
          legend columns=3,
          legend style={font=\scriptsize, draw=none, fill=none,
                        /tikz/every even column/.append style={column sep=6pt}},
        ]
        \addplot+[hy3 lmetric] coordinates {(TTFT,63.47853) (Queueing,57.84457)};
        \addlegendentry{LMetric}
        \addplot+[hy3 smetric] coordinates {(TTFT,1.55946) (Queueing,0.36244)};
        \addlegendentry{SMetric}
        \addplot+[hy3 vesta] coordinates {(TTFT,1.33291) (Queueing,0.14739)};
        \addlegendentry{Vesta}
      \end{axis}
      \begin{axis}[
          hy3 mean,
          hy3 break upper,
          at={([yshift=2.5pt]hy3meanlow.north west)},
          ymin=45, ymax=68,
          ytick={50,65},
        ]
        \addplot+[hy3 lmetric] coordinates {(TTFT,63.47853) (Queueing,57.84457)};
        \addplot[draw=none, fill=none] coordinates {(TTFT,35) (Queueing,35)};
        \addplot[draw=none, fill=none] coordinates {(TTFT,35) (Queueing,35)};
      \end{axis}
      \hybreakmark{hy3meanlow}
    \end{tikzpicture}
    \caption{Mean TTFT and queueing delay.}
    \label{fig:sota-hy3-mean}
  \end{subfigure}\hfill
  \begin{subfigure}[t]{0.48\columnwidth}
    \centering
    \begin{tikzpicture}
      \begin{axis}[
          hy3 axis,
          hy3 values,
          ybar,
          bar width=12pt,
          bar shift=0pt,
          xmin=0.4, xmax=3.6,
          xtick={1,2,3},
          xticklabels={LMetric,SMetric,Vesta},
          ymax=100,
          ytick={0,25,50,75,100},
          ylabel={KV cache hit rate (\%)},
        ]
        \addplot+[hy3 lmetric] coordinates {(1,61.18)};
        \addplot+[hy3 smetric] coordinates {(2,91.67)};
        \addplot+[hy3 vesta] coordinates {(3,91.13)};
      \end{axis}
    \end{tikzpicture}
    \caption{KV cache hit rate.}
    \label{fig:sota-hy3-hit}
  \end{subfigure}\par\smallskip
  \begin{subfigure}[t]{0.48\columnwidth}
    \centering
    \begin{tikzpicture}
      \begin{axis}[
          hy3 tail,
          hy3 break lower,
          name=hy3ttftlow,
          ymax=22,
          ytick={0,5,10,15,20},
          ylabel={TTFT (s)},
          ylabel style={at={(ticklabel cs:0.75)}},
        ]
        \addplot+[hy3 lmetric] coordinates {
          (P50,66.288) (P90,117.0049) (P95,118.4703) (P99,119.657)};
        \addplot+[hy3 smetric] coordinates {
          (P50,0.399) (P90,3.4266) (P95,6.98580) (P99,20.79936)};
        \addplot+[hy3 vesta] coordinates {
          (P50,0.401) (P90,3.1712) (P95,5.8568) (P99,14.76462)};
      \end{axis}
      \begin{axis}[
          hy3 tail,
          hy3 break upper,
          at={([yshift=2.5pt]hy3ttftlow.north west)},
          ymin=30, ymax=126,
          ytick={60,90,120},
        ]
        \addplot+[hy3 lmetric] coordinates {
          (P50,66.288) (P90,117.0049) (P95,118.4703) (P99,119.657)};
        \addplot[draw=none, fill=none] coordinates {
          (P50,30) (P90,30) (P95,30) (P99,30)};
        \addplot[draw=none, fill=none] coordinates {
          (P50,30) (P90,30) (P95,30) (P99,30)};
      \end{axis}
      \hybreakmark{hy3ttftlow}
    \end{tikzpicture}
    \caption{TTFT percentiles.}
    \label{fig:sota-hy3-ttft}
  \end{subfigure}\hfill
  \begin{subfigure}[t]{0.48\columnwidth}
    \centering
    \begin{tikzpicture}
      \begin{axis}[
          hy3 tail,
          hy3 break lower,
          name=hy3queuelow,
          ymax=12,
          ytick={0,4,8},
          ylabel={Queueing delay (s)},
          ylabel style={at={(ticklabel cs:0.75)}},
        ]
        \addplot+[hy3 lmetric] coordinates {
          (P50,59.919) (P90,111.197) (P95,113.729) (P99,116.511)};
        \addplot+[hy3 smetric] coordinates {
          (P50,0) (P90,0) (P95,1.1024) (P99,10.70696)};
        \addplot+[hy3 vesta] coordinates {
          (P50,0) (P90,0) (P95,0.002) (P99,4.33014)};
      \end{axis}
      \begin{axis}[
          hy3 tail,
          hy3 break upper,
          at={([yshift=2.5pt]hy3queuelow.north west)},
          ymin=30, ymax=120,
          ytick={60,90,120},
        ]
        \addplot+[hy3 lmetric] coordinates {
          (P50,59.919) (P90,111.197) (P95,113.729) (P99,116.511)};
        \addplot[draw=none, fill=none] coordinates {
          (P50,30) (P90,30) (P95,30) (P99,30)};
        \addplot[draw=none, fill=none] coordinates {
          (P50,30) (P90,30) (P95,30) (P99,30)};
      \end{axis}
      \hybreakmark{hy3queuelow}
    \end{tikzpicture}
    \caption{Queueing-delay percentiles.}
    \label{fig:sota-hy3-queue}
  \end{subfigure}
  \caption{Serving performance on Hy3-FP8 with 6P6D.}
  \Description{Four bar charts compare LMetric, SMetric, and Vesta on Hy3-FP8
  with six prefill and six decode vLLM instances. Vesta has the lowest mean
  TTFT and the lowest mean queueing delay, and the lowest TTFT at P90, P95,
  and P99. Its cache
  hit rate is 91.13 percent, compared with 91.67 percent for SMetric and
  61.18 percent for LMetric. SMetric and Vesta have zero queueing delay at
  P50 and P90; at P95 and P99 Vesta queues far less than SMetric. The y axes of
  the first, third, and fourth charts are broken so that the much smaller
  SMetric and Vesta bars stay readable.}
  \label{fig:sota-compare-hy3-6p6d}
\end{figure}
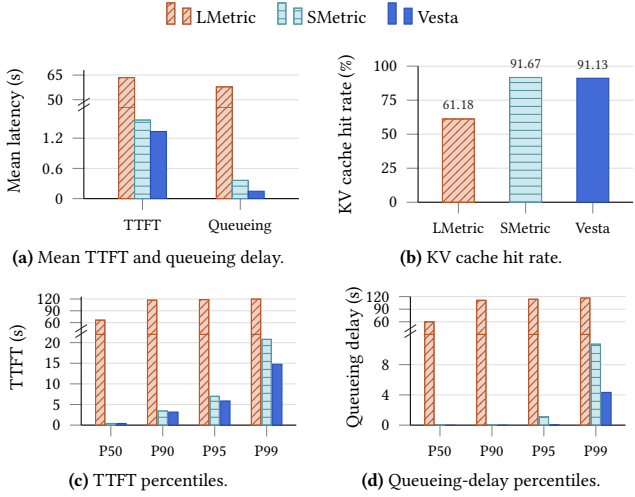

\begin{figure}[!tb]
  \centering
  \includegraphics[width=0.74\columnwidth]{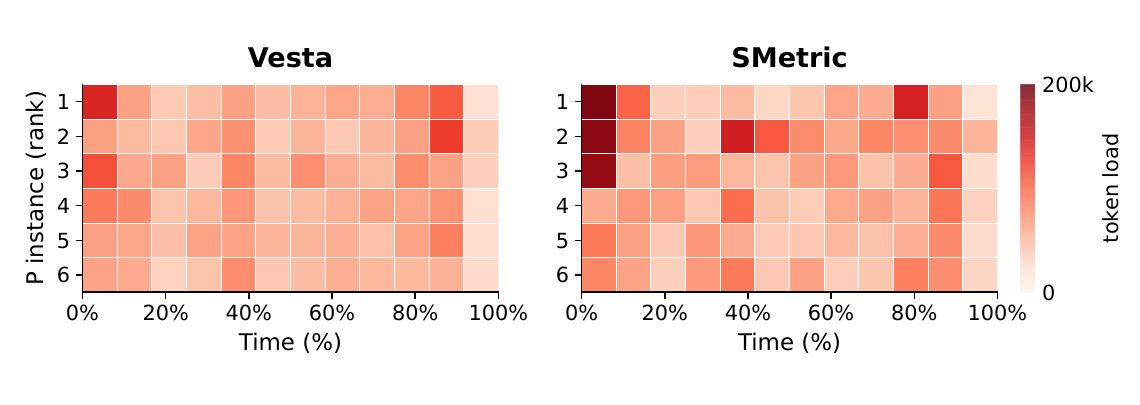}
  \caption{Hy3-FP8 prefill token load with 6P6D.}
  \Description{Vesta (left) and SMetric (right) heatmaps for six Hy3-FP8
  prefill instances. The mean max-to-mean ratio is 1.521 for SMetric and
  1.289 for Vesta. All 12 time bins are fully observed.}
  \label{fig:prefill-load-hy3}
\end{figure}

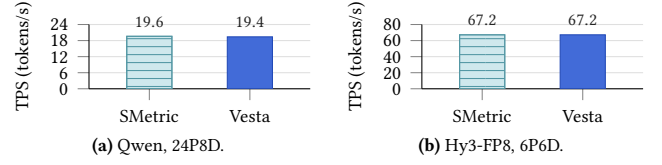
\begin{figure}[!tb]
  \centering
  \captionsetup[subfigure]{font=scriptsize,skip=2pt}
  \pgfplotsset{
    pd tps axis/.style={
      scale only axis,
      width=0.72\linewidth,
      height=0.85cm,
      ybar,
      bar width=17pt,
      bar shift=0pt,
      xmin=0.4, xmax=2.6,
      xtick={1,2},
      xticklabels={SMetric,Vesta},
      ymin=0,
      axis lines*=left,
      ymajorgrids=true,
      grid style={draw=vestaGray!30},
      tick align=outside,
      tick label style={font=\scriptsize},
      label style={font=\scriptsize},
      scaled y ticks=false,
      ylabel={TPS (tokens/s)},
      nodes near coords,
      every node near coord/.append style={
        font=\fontsize{6}{7}\selectfont,
        text=vestaDark,
        /pgf/number format/fixed,
        /pgf/number format/precision=1,
        /pgf/number format/fixed zerofill,
      },
    },
    pd tps smetric/.style={
      draw=vestaCyan!80!vestaDark,
      fill=vestaCyan!30,
      postaction={pattern=horizontal lines, pattern color=vestaCyan!80!vestaDark},
    },
    pd tps vesta/.style={draw=vestaBlue!85!vestaDark, fill=vestaBlue!85},
  }
  \begin{subfigure}[t]{0.48\columnwidth}
    \centering
    \begin{tikzpicture}
      \begin{axis}[pd tps axis, ymax=24, ytick={0,6,12,18,24}]
        \addplot+[pd tps smetric] coordinates {(1,19.6)};
        \addplot+[pd tps vesta] coordinates {(2,19.4)};
      \end{axis}
    \end{tikzpicture}
    \caption{Qwen, 24P8D.}
    \label{fig:pd-tps-qwen}
  \end{subfigure}\hfill
  \begin{subfigure}[t]{0.48\columnwidth}
    \centering
    \begin{tikzpicture}
      \begin{axis}[pd tps axis, ymax=80, ytick={0,20,40,60,80}]
        \addplot+[pd tps smetric] coordinates {(1,67.2)};
        \addplot+[pd tps vesta] coordinates {(2,67.2)};
      \end{axis}
    \end{tikzpicture}
    \caption{Hy3-FP8, 6P6D.}
    \label{fig:pd-tps-hy3}
  \end{subfigure}
  \caption{TPS under PD disaggregation: (a) Qwen2.5-7B-Instruct-1M,
  24P8D; (b) Hy3-FP8, 6P6D.}
  \Description{Two bar charts compare SMetric and Vesta at fixed deployments.
  Qwen TPS is 19.668 for SMetric and 19.448 for Vesta. Hy3-FP8 TPS is
  67.207 for SMetric and 67.213 for Vesta. TPS is measured in tokens per second.}
  \label{fig:pd-tps-comparison}
\end{figure}

\noindent\textbf{Hy3-FP8.}
We next evaluate Hy3-FP8 on a 6P6D deployment with 644 sessions (11K requests) sampled from \emph{Trace Agent}.

\Cref{fig:sota-compare-hy3-6p6d} confirms the same pattern on Hy3-FP8.
Vesta lowers mean TTFT to 1,332.91~ms, $14.5\%$ below SMetric and its cache hit rate remains close to SMetric's.  The gain again comes from
queueing: mean queueing delay falls from 362.44 to
147.39~ms, which accounts for $94.9\%$ of the mean TTFT
reduction.  The improvement grows toward the tail: median TTFT
is unchanged at $0.40$~s, while P90, P95, and P99 TTFT fall by $7.5\%$,
$16.2\%$, and $29.0\%$.  This mirrors where the two systems queue at all:
neither waits at P50 or P90, and the gap widens from 1.10~s at P95 to
6.38~s at P99.

\noindent\textbf{Prefill load balance.}
\Cref{fig:prefill-load-hy3} compares the prefill token load of SMetric and Vesta
on Hy3-FP8.  Vesta again weakens
SMetric's hotspots, lowering the maximum instance-bin load from 190.0k to 140.1k
tokens ($26.3\%$).  Over all 12 fully observed bins, the mean max-to-mean ratio
falls from $1.521\times$ to $1.289\times$ ($15.2\%$).  Unlike on Qwen, where
Vesta mainly reduces absolute hotspot intensity, here it also balances load
markedly more evenly across the six prefill instances.

\noindent\textbf{TPS.}
\Cref{fig:pd-tps-comparison} shows that Vesta retains essentially the same
TPS as SMetric: 19.4 versus 19.6~tokens/s on Qwen and 67.2~tokens/s on Hy3-FP8.  Thus, the TTFT reductions above do not come at
the expense of throughput.

\subsection{Ablation Study}
\paragraph{Comparison with Vesta Variants.}
In this experiment, we evaluate the impact of the congestion-priced completion-time objective and the write-back protocol on the serving performance.
We evaluate Qwen2.5-7B-Instruct-1M on a 6P2D deployment by replaying 8,030
requests from 453 sessions sampled from \emph{Trace Agent}, emulating
peak-period demand at an average rate of 2.07~requests/s.

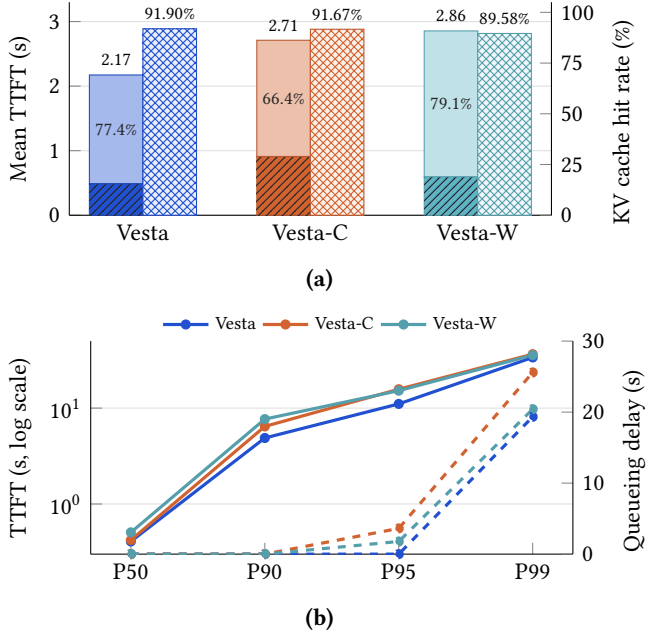
\begin{figure}[!tb]
  \centering
  \begin{subfigure}[t]{\columnwidth}
    \centering
    \begin{tikzpicture}
      \begin{axis}[
          name=meanAxis,
          width=0.94\linewidth,
          height=4.4cm,
          xmin=0.4, xmax=7.6,
          xtick={1.5,4,6.5},
          xticklabels={Vesta,Vesta-C,Vesta-W},
          ymin=0, ymax=3.3,
          ytick={0,1,2,3},
          ylabel={Mean TTFT (s)},
          ylabel style={font=\small},
          xticklabel style={font=\small},
          yticklabel style={font=\small},
          axis lines*=left,
          ymajorgrids=true,
          grid style={draw=vestaGray!30},
          clip=false,
        ]
        \path[draw=vestaBlue!85!vestaDark, fill=vestaBlue,
              postaction={pattern=north east lines, pattern color=vestaDark}]
          (axis cs:0.70,0) rectangle (axis cs:1.50,0.49081);
        \path[draw=vestaBlue!85!vestaDark, fill=vestaBlue!40]
          (axis cs:0.70,0.49081) rectangle (axis cs:1.50,2.17275);
        \path[draw=vestaOrange!85!vestaDark, fill=vestaOrange,
              postaction={pattern=north east lines, pattern color=vestaDark}]
          (axis cs:3.20,0) rectangle (axis cs:4.00,0.91089);
        \path[draw=vestaOrange!85!vestaDark, fill=vestaOrange!40]
          (axis cs:3.20,0.91089) rectangle (axis cs:4.00,2.71055);
        \path[draw=vestaCyan!85!vestaDark, fill=vestaCyan,
              postaction={pattern=north east lines, pattern color=vestaDark}]
          (axis cs:5.70,0) rectangle (axis cs:6.50,0.59682);
        \path[draw=vestaCyan!85!vestaDark, fill=vestaCyan!40]
          (axis cs:5.70,0.59682) rectangle (axis cs:6.50,2.85503);
        \coordinate (meanVesta) at (axis cs:1.10,2.17275);
        \coordinate (meanVestaC) at (axis cs:3.60,2.71055);
        \coordinate (meanVestaW) at (axis cs:6.10,2.85503);
        \coordinate (computeVesta) at (axis cs:1.10,1.33178);
        \coordinate (computeVestaC) at (axis cs:3.60,1.81072);
        \coordinate (computeVestaW) at (axis cs:6.10,1.72593);
      \end{axis}
      \begin{axis}[
          at={(meanAxis.south west)}, anchor=south west,
          width=0.94\linewidth,
          height=4.4cm,
          xmin=0.4, xmax=7.6,
          xtick=\empty,
          ymin=0, ymax=105,
          ytick={0,25,50,75,100},
          ylabel={KV cache hit rate (\%)},
          ylabel style={font=\small, xshift=-5pt},
          yticklabel style={font=\small},
          axis x line=none,
          axis y line*=right,
          ymajorgrids=false,
          clip=false,
        ]
        \path[draw=vestaBlue!85!vestaDark, fill=vestaBlue!10,
              postaction={pattern=crosshatch, pattern color=vestaBlue}]
          (axis cs:1.50,0) rectangle (axis cs:2.30,91.90);
        \path[draw=vestaOrange!85!vestaDark, fill=vestaOrange!10,
              postaction={pattern=crosshatch, pattern color=vestaOrange}]
          (axis cs:4.00,0) rectangle (axis cs:4.80,91.67);
        \path[draw=vestaCyan!85!vestaDark, fill=vestaCyan!10,
              postaction={pattern=crosshatch, pattern color=vestaCyan!85!vestaDark}]
          (axis cs:6.50,0) rectangle (axis cs:7.30,89.58);
        \coordinate (hitVesta) at (axis cs:1.90,91.90);
        \coordinate (hitVestaC) at (axis cs:4.40,91.67);
        \coordinate (hitVestaW) at (axis cs:6.90,89.58);
      \end{axis}
      \node[font=\scriptsize, anchor=south] at (meanVesta) {2.17};
      \node[font=\scriptsize, anchor=south] at (meanVestaC) {2.71};
      \node[font=\scriptsize, anchor=south] at (meanVestaW) {2.86};
      \node[font=\scriptsize, text=vestaDark] at (computeVesta) {77.4\%};
      \node[font=\scriptsize, text=vestaDark] at (computeVestaC) {66.4\%};
      \node[font=\scriptsize, text=vestaDark] at (computeVestaW) {79.1\%};
      \node[font=\scriptsize, anchor=south] at (hitVesta) {91.90\%};
      \node[font=\scriptsize, anchor=south] at (hitVestaC) {91.67\%};
      \node[font=\scriptsize, anchor=south] at (hitVestaW) {89.58\%};
    \end{tikzpicture}
    \caption{}
    \label{fig:variant-mean-ttft}
  \end{subfigure}\par\medskip
  \begin{subfigure}[t]{\columnwidth}
    \centering
    \begin{tikzpicture}
      \begin{axis}[
          name=ttftAxis,
          width=0.94\linewidth,
          height=4.4cm,
          symbolic x coords={P50,P90,P95,P99},
          xtick=data,
          ymin=0.3, ymax=50,
          ymode=log,
          log basis y=10,
          ytick={1,10},
          ylabel={TTFT (s, log scale)},
          ylabel style={font=\small},
          xticklabel style={font=\small},
          yticklabel style={font=\small},
          axis lines*=left,
          ymajorgrids=true,
          grid style={draw=vestaGray!30},
          legend columns=3,
          legend style={font=\scriptsize, at={(0.5,1.00)}, anchor=south,
                        draw=none, fill=none, inner sep=1pt},
        ]
        \addplot[draw=vestaBlue, very thick, mark=*, mark size=1.3pt,
                 mark options={draw=vestaBlue, fill=vestaBlue}]
          coordinates {(P50,0.407) (P90,4.88040) (P95,11.05935) (P99,33.76155)};
        \addlegendentry{Vesta}
        \addplot[draw=vestaOrange, very thick, mark=*, mark size=1.3pt,
                 mark options={draw=vestaOrange, fill=vestaOrange}]
          coordinates {(P50,0.421) (P90,6.44240) (P95,15.73700) (P99,36.51388)};
        \addlegendentry{Vesta-C}
        \addplot[draw=vestaCyan!85!vestaDark, very thick, mark=*, mark size=1.3pt,
                 mark options={draw=vestaCyan!85!vestaDark, fill=vestaCyan!85!vestaDark}]
          coordinates {(P50,0.506) (P90,7.68230) (P95,15.19950) (P99,35.67539)};
        \addlegendentry{Vesta-W}
      \end{axis}
      \begin{axis}[
          at={(ttftAxis.south west)},
          anchor=south west,
          width=0.94\linewidth,
          height=4.4cm,
          symbolic x coords={P50,P90,P95,P99},
          xtick=\empty,
          ymin=0, ymax=30,
          ytick={0,10,20,30},
          ylabel={Queueing delay (s)},
          ylabel style={font=\small, xshift=-5pt},
          yticklabel style={font=\small},
          axis x line=none,
          axis y line*=right,
          ymajorgrids=false,
        ]
        \addplot[draw=vestaBlue, very thick, dashed, mark=*, mark size=1.3pt,
                 mark options={draw=vestaBlue, fill=vestaBlue}]
          coordinates {(P50,0) (P90,0) (P95,0) (P99,19.34326)};
        \addplot[draw=vestaOrange, very thick, dashed, mark=*, mark size=1.3pt,
                 mark options={draw=vestaOrange, fill=vestaOrange}]
          coordinates {(P50,0) (P90,0) (P95,3.62180) (P99,25.58452)};
        \addplot[draw=vestaCyan!85!vestaDark, very thick, dashed, mark=*, mark size=1.3pt,
                 mark options={draw=vestaCyan!85!vestaDark, fill=vestaCyan!85!vestaDark}]
          coordinates {(P50,0) (P90,0) (P95,1.78075) (P99,20.41791)};
      \end{axis}
    \end{tikzpicture}
    \caption{}
    \label{fig:variant-tail-latency}
  \end{subfigure}
  \caption{Latency comparison among Vesta variants. (a) Mean TTFT
  decomposition and KV cache hit rate; (b) TTFT (solid) and queueing-delay
  (dashed) percentiles.}
  \label{fig:vesta-variant-latency}
\end{figure}

We construct \emph{Vesta-C} by removing the final, congestion-pricing term from~\Cref{eq:executor-objective} to isolate the contribution of accounting for the queueing externality.
In \emph{Vesta-W}, a remote executor offloads the KV cache
to its L2 cache when it fetches the remote cache, and the newly generated KV cache is retained and offloaded to local L2 cache
rather than written back to the original holder.  Consequently, the remote
executor becomes the session's new holder.  Unlike Vesta, which determines a
stable holder by hashing the session ID, \emph{Vesta-W} selects the holder for
each request according to the instance with the longest matched prefix; thus,
each migration can change the session's holder.

\Cref{fig:vesta-variant-latency} compares the Vesta variants.  Compared with
\emph{Vesta-C}, Vesta reduces mean TTFT by $19.8\%$ (from $2.71$ to $2.17$~s)
and mean queueing delay by $46.1\%$ (from $0.91$ to $0.49$~s), while preserving
an essentially identical KV cache hit rate ($91.90\%$ versus $91.67\%$).  It
also reduces P95 and P99 TTFT by $29.7\%$ and $7.5\%$, respectively.  These
improvements show that accounting for the queueing externality prevents an
executor choice that appears locally attractive from imposing disproportionate
queueing delay on subsequent requests.

Vesta also outperforms \emph{Vesta-W}, reducing mean TTFT by $23.9\%$ (from
$2.86$ to $2.17$~s) and mean queueing delay by $17.8\%$ (from $0.60$ to
$0.49$~s).  \emph{Vesta-W} changes a session's holder after a migration,
causing the same session's KV cache to be replicated across L2 caches.  These
replicas reduce effective cache capacity and increase cache eviction, lowering
the KV cache hit rate by $2.32$ percentage points (from $91.90\%$ to
$89.58\%$).  The resulting loss of reuse increases the fraction of TTFT spent in
computation from $77.4\%$ to $79.1\%$ and raises mean computation time from
$1.68$ to $2.26$~s.  Maintaining a stable session holder and writing newly
generated KV cache back to it is therefore essential for preserving cache
capacity and long-term reuse.

\paragraph{Cross-Bucket Scheduling.}
We evaluate Vesta's cross-bucket scheduling by replaying 401 sessions (15.7K requests) sampled from \emph{Trace Agent} on
Qwen2.5-7B-Instruct-1M.  We deploy a 6P2D cluster for the 32K bucket and an
18P6D cluster for the 256K bucket.  With cross-bucket migration, when a session
outgrows the 32K bucket, Vesta transfers its longest reusable prefix from the
source-bucket holder to the session's hashed holder in the 256K bucket.  The
target holder retains the transferred prefix and subsequently appends the new KV
cache generated in the larger bucket.  Without migration, the target-bucket
holder has no inherited prefix and must prefill the session context from
scratch on its first request in the 256K bucket.

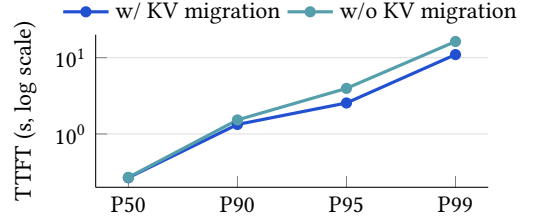
\begin{figure}[!tb]
  \centering
  \begin{tikzpicture}
    \begin{axis}[
        width=0.8\columnwidth,
        height=3.6cm,
        symbolic x coords={P50,P90,P95,P99},
        xtick=data,
        ymin=0.2, ymax=20,
        ymode=log,
        log basis y=10,
        ytick={1,10},
        ylabel={TTFT (s, log scale)},
        ylabel style={font=\small},
        xticklabel style={font=\small},
        yticklabel style={font=\small},
        axis lines*=left,
        ymajorgrids=true,
        grid style={draw=vestaGray!30},
        legend columns=2,
        legend style={font=\small, at={(0.5,1.03)}, anchor=south,
                      draw=none, fill=none, inner sep=1pt},
      ]
      \addplot[draw=vestaBlue, very thick, mark=*, mark size=1.5pt,
               mark options={draw=vestaBlue, fill=vestaBlue}]
        coordinates {(P50,0.267) (P90,1.332) (P95,2.54880) (P99,10.97592)};
      \addlegendentry{w/ KV migration}
      \addplot[draw=vestaCyan!85!vestaDark, very thick, mark=*, mark size=1.5pt,
               mark options={draw=vestaCyan!85!vestaDark, fill=vestaCyan!85!vestaDark}]
        coordinates {(P50,0.270) (P90,1.52760) (P95,3.95780) (P99,16.29472)};
      \addlegendentry{w/o KV migration}
    \end{axis}
  \end{tikzpicture}
  \caption{Impact of cross-bucket migration on TTFT in the 256K bucket.  Migration substantially reduces tail TTFT.}
  \label{fig:cross-bucket-migration}
\end{figure}

\Cref{fig:cross-bucket-migration} shows that cross-bucket migration leaves
median TTFT essentially unchanged ($0.267$ versus $0.270$~s), but substantially
improves tail latency.  Relative to recomputing the prefix after a bucket
transition, migration reduces P90, P95, and P99 TTFT by $12.8\%$, $35.6\%$, and
$32.6\%$, respectively.  These gains show that preserving a session's reusable
KV cache across bucket boundaries avoids costly cold starts precisely for the
long-context requests that dominate tail latency.

\section{Related Work}
\label{sec:related}

\subsection{LLM Scheduling}
At the cluster level, online schedulers adapt request placement to changing
load.  Llumnix performs cross-instance rescheduling through live migration of
in-memory request state, whereas AdaGen adapts cluster routing and compute
layouts to workload changes~\cite{osdi24_llumnix,eurosys26_adagen}.
Recent work combines cache-locality and load signals in request
placement~\cite{osdi26_lmetric,iclr26_dualmap}, and SMetric adopts a
session-centric policy that load-balances the first request of a session before
favoring cache affinity for its follow-up requests~\cite{arxiv26_smetric}.
Analytical schedulers further estimate candidate-instance latency or SLO
compliance to guide placement~\cite{arxiv25_slosserve,arxiv25_polyserve,fast25_mooncake}.
Vesta likewise reconciles cache reuse with load balance, but from a
queueing-theoretic perspective: it decouples stable session cache ownership from
per-request execution and weighs the queueing benefit of migration against its
cache-transfer cost under the current cluster load.

\subsection{LLM Inference Optimization}


\noindent\textbf{Token and KV Cache Compression.}
Long contexts make attention and KV cache storage the dominant costs of
inference.  Sparse-attention and token-selection methods select a small set of context
tokens for each query, reducing the attention work during prefill and decode
~\cite{nips24_minference,iclr2025_flexprefill,streamingllm,snapkv}.
KV cache quantization instead stores keys and values at lower precision,
reducing the memory footprint and allowing a fixed-capacity cache tier to
retain more prefixes~\cite{kivi,kvquant}.

\noindent\textbf{Model Architecture.}
Architectural co-design changes the representation or computation performed by
the model itself.  Linear-attention architectures replace the quadratic
softmax-attention operation with a recurrent or kernelized formulation whose
cost scales linearly with context length~\cite{katharopoulos2020transformers,choromanski2020rethinking,yang2023gated}.
Grouped-query attention reduces the number of distinct KV heads during
decoding~\cite{gqa}, while recent long-context models combine multi-head latent
attention with sparse-attention mechanisms to reduce KV cache size and
attention cost~\cite{arxiv25_deepseekv3.2,arxiv26_deepseekv4}.
These techniques improve the execution efficiency of an individual request and are
complementary to Vesta, which instead operates at the cluster level.

\section{Conclusion}
\label{sec:conclusion}

We presented Vesta, a scheduler that grounds agentic LLM serving in queueing
theory.  Vesta maintains a stable holder for each session's KV cache while
selecting an executor per request, using spare capacity without scattering that
cache.  Its objective prices the queueing externality of a placement, migrating
execution under slack and favoring the least additional transfer and prefill
work as congestion rises.  Vesta also migrates reusable prefixes across length
buckets to avoid cold starts as sessions lengthen.  On production agentic traces,
Vesta reduces mean and tail TTFT in both serving architectures while achieving
higher or comparable throughput.

\ifnum0=1
\begin{acks}
  \TODO{Camera-ready only.}
\end{acks}
\fi
\bibliographystyle{ACM-Reference-Format}
\bibliography{ref/Top,ref/all}

\end{document}